\documentclass[11pt]{article}

\usepackage{a4wide}
\usepackage{xspace}
\usepackage{multicol}
\usepackage{appendix}
\usepackage{amsmath}
\usepackage{amssymb}
\usepackage{jheppub} 

\usepackage{graphicx}
\usepackage{epsfig}
\usepackage{comment}
\usepackage{epsfig}
\usepackage{tikz}
\usetikzlibrary{decorations.pathmorphing}
\usetikzlibrary{decorations.markings}
\usetikzlibrary{positioning, shapes, snakes, arrows}
\tikzset{
	quark/.style={postaction={decorate},
		decoration={markings,mark=at position .5 with {\arrow[#1]{latex}}}},
	scalar/.style={dashed,postaction={decorate},
		decoration={markings,mark=at position .5 with {\arrow[#1]{latex}}}},
	gluon/.style={decorate,
		decoration={coil,amplitude=2pt, segment length=2pt,  pre length=.1cm, post length=.1cm}},
	boson/.style={-latex,decorate, decoration={snake, segment length=4pt, amplitude=1.8pt, pre length=.1cm, post length=.25cm}},
	photon/.style={decorate, decoration={snake, segment length=4pt, amplitude=1.8pt,  pre length=.1cm, post length=.1cm}},
	dphoton/.style={decorate, decoration={snake, segment length=4pt, amplitude=1.8pt,  pre length=.1cm, post length=.25cm},-latex}
}

\usepackage{hyperref} 
\hypersetup{linktocpage} 
\hypersetup{
    colorlinks,
    citecolor=blue,
    filecolor=blue,
    linkcolor=blue,
    urlcolor=blue}
\usepackage{amssymb}

\newcommand{\beq}{\begin{equation}}
\newcommand{\eeq}{\end{equation}}

\newcommand*\pFq[2]{{}_{#1}F_{#2}}

\newcommand{\bdm}{\begin{displaymath}}
\newcommand{\edm}{\end{displaymath}}

\newcommand\f[2]{\frac{#1}{#2}}
\def\Ph{{\hat P}}

\def\beeq{\begin{eqnarray}}
\def\eeeq{\end{eqnarray}}
\def\ep{\epsilon}
\def\nn{\nonumber}
\def\lra{\leftrightarrow}

\newcommand{\bea}{\begin{eqnarray}}
\newcommand{\eea}{\end{eqnarray}}
\newcommand{\la}{\left\langle}
\newcommand{\ra}{\right\rangle}

\usepackage{color}
\definecolor{darkgreen}{rgb}{0,0.5,0}
\definecolor{darkblue}{rgb}{0,0,0.7}
\definecolor{darkred}{rgb}{0.5,0,0.0}
\definecolor{darkorange}{rgb}{0.8,0.4,0.0}

\def\d{\partial}

\def \d{{\rm d} }
\def \d0 {D\O \;}

\def \zc{z_\text{cut}}
\def \event2{\textsc{Event2} }

\makeatletter
\g@addto@macro\bfseries{\boldmath}
\makeatother
\title{\textbf{Dissecting the collinear structure of quark splitting at NNLL}}
\author{Mrinal Dasgupta and Basem Kamal El-Menoufi}
\affiliation{Lancaster-Manchester-Sheffield Consortium for Fundamental Physics, School of Physics
  \& Astronomy, University of Manchester, Manchester M13 9PL, United Kingdom}
  \emailAdd{mrinal.dasgupta@manchester.ac.uk, basem.el-menoufi@manchester.ac.uk}
\abstract{We explore the collinear limit of final-state quark splittings at order $\alpha_s^2$. While at general NLL level, this limit is described simply by a product of leading-order $1\to 2$ DGLAP splitting functions, at the NNLL level we need to consider $1\to3$ splitting functions. Here, by performing suitable integrals of the triple-collinear splitting functions, we demonstrate how one may extract $\mathcal{B}^q_2(z)$, a differential version of the coefficient $B^q_2$ that enters the quark form factor at NNLL and governs the intensity of collinear radiation from a quark. The variable $z$ corresponds to the quark energy fraction after an initial $1 \to 2$ splitting, and our results yield effective higher-order splitting functions, which may be considered  as a step towards the construction of NNLL parton showers.  
Further, while in the limit $z \to 1$ we recover the standard soft limit results involving the CMW coupling with scale $k_t$, the $z$ dependence we obtain also motivates the extension of the notion of a physical coupling beyond the soft limit.}

\begin{document}

\maketitle

\section{Introduction}
QCD has long been confirmed as the theory of strong interactions and is now decisively in the precision era. The demand for percent-level precision on QCD calculations is primarily being driven by LHC phenomenology with the large volume of experimental data accumulated at the LHC and the focus on firmly establishing the properties of the Standard Model Higgs sector as well as the search for possible subtle signs of new physics \cite{KeithEllis:2019bfl,Azzi:2019yne,Cepeda:2019klc,CidVidal:2018eel}. 

On the theoretical QCD side, in the context of precision, there has been particularly impressive recent progress in fixed-order perturbative calculations (see Ref.~\cite{Heinrich:2020ybq} for a review and further references). For a number of LHC applications however, one often needs to go beyond fixed-order calculations which are perturbative approximations involving a small number of partons, in contrast to the high-multiplicity hadronic final states seen in practice  at colliders. The need for predictions beyond fixed-order becomes particularly obvious when one encounters observables sensitive to widely disparate scales and hence large logarithms in scale ratios, which require resummation. 

Resummed calculations and the techniques behind them have also undergone substantial development over the past couple of decades.  For a broad class of observables that have a property called recursive infrared and collinear (rIRC) safety \cite{Banfi:2004yd} we can write resummed results for the cumulative distribution, i.e the distribution integrated up to some value $v=e^{-L}$,
in a standard exponentiated form \cite{Catani:1991kz}: 
\begin{equation}
\Sigma(\alpha_s,\alpha_s L) = \exp \left[\alpha_s^{-1}g_1(\alpha_s L)+g_2(\alpha_sL) +\alpha_s g_3(\alpha_s L)+\cdots \right],
\end{equation}
where knowledge of the functions $g_1$, $g_2$ and  $g_3$ corresponds to achieving leading logarithmic (LL), next-to--leading logarithmic (NLL) and next-to--next-to leading (NNLL) logarithmic accuracy respectively. The  state-of--the art has  progressed from the NLL accuracy achievable for a limited number of observables in the early 1990's to achievement of automated NLL calculations for a wide range of global observables \cite{Banfi:2001bz,Banfi:2004yd,Banfi:2004nk}, and the development of NNLL (and in some cases essentially  $\mathrm{N^3LL}$ ) resummed calculations for the main classes of resummation \cite{Bozzi:2003jy,Bozzi:2005wk,Becher:2008cf,Berger:2010xi,Abbate:2010xh,Becher:2010tm,Stewart:2010pd,Banfi:2011dx,Jouttenus:2011wh,Zhu:2012ts,Becher:2012qa,Becher:2012qc,Banfi:2012jm,Stewart:2013faa,Becher:2013xia,Kang:2013lga,Kang:2013wca,Kang:2013nha,Hoang:2014wka,Banfi:2014sua,Becher:2014aya,Catani:2014qha,Becher:2015lmy,Frye:2016okc,Banfi:2016zlc,Tulipant:2017ybb,Bizon:2017rah,Bizon:2018foh,Procura:2018zpn,Moult:2018jzp,Bell:2018mkk,Chen:2018pzu,Banfi:2018mcq}, including progress on automated resummation frameworks  \cite{Banfi:2014sua,Banfi:2018mcq,Becher:2014aya,Bell:2018mkk}. However it remains true that the scope of analytic (or semi-numerical) resummation is still limited to a few types of observable which can be easily understood analytically in specific limits. On the other hand the range of observables that is relevant for phenomenology is ever increasing. Some notable examples come from the field of jet substructure in the boosted domain, where for the purposes of tagging and grooming it is common to simultaneously cut on a range of variables in order to achieve good discrimination between signal and background. In such cases, while resummation is essential to describe the results obtained, it is often hard to achieve analytically beyond the most basic modified LL accuracy (see Ref.~\cite{Dasgupta:2021kgi} for a recent example from top tagging).

Given the limitations of current analytic resummation approaches, it is important to consider other available tools for handling general multiscale problems. This brings us to parton shower algorithms which are at the core of general purpose Monte Carlo event  generators programs \cite{Buckley:2011ms}. These, by their very nature, provide all-order perturbative results encapsulating the resummation of large logarithms for general observables. Parton showers have very broad applicability, but the question of understanding their logarithmic accuracy has proved elusive until recently. While showers are often stated to achieve leading logarithmic accuracy, and also include the main ingredients necessary for reaching NLL accuracy, Ref.~\cite{Dasgupta:2018nvj} demonstrated that a widely used class of dipole showers, including the Pythia transverse-momentum ordered shower \cite{Sjostrand:2004ef}, which is the default shower in the Pythia8 program \cite{Sjostrand:2014zea}, failed to achieve general LL accuracy beyond the leading colour approximation and also failed to reach NLL accuracy for a number of common observables even at leading colour. Angular-ordered showers on the other hand have long been known to fail NLL accuracy criteria for non-global observables \cite{Banfi:2006gy,Dasgupta:2001sh}. With a better understanding emerging of the limitations of various classes of showers, it has been possible to identify a set of principles that should be satisfied for a shower to be deemed NLL accurate \cite{Dasgupta:2020fwr}. NLL accurate showers, the PanScales showers,  based on these principles have been constructed and numerically demonstrated to achieve full NLL accuracy for a wide range of global observables, including also subleading colour effects and spin correlations \cite{Dasgupta:2020fwr,Hamilton:2020rcu,Karlberg:2021kwr}.  Other candidates for NLL accurate showers have also been proposed and analytically demonstrated, to achieve NLL accuracy for the thrust \cite{Forshaw:2020wrq,Holguin:2020joq,Nagy:2020dvz,Nagy:2020rmk} and the jet multiplicity \cite{Forshaw:2020wrq,Holguin:2020joq}.

Given the recent spurt in activity in the context of NLL showers, which demonstrates that showers can be designed to achieve broad NLL accuracy, it is legitimate to think about whether further advances in resummation can be brought to bear on the construction of NNLL accurate showers. This is a substantially more challenging problem, but solutions to it would be invaluable in the context of extending theoretical precision for collider studies. A first obvious step in thinking about NNLL showers would be to consider the higher-order ingredients that shall be needed in order to extend shower accuracy. Once identified, the inclusion of these ingredients consistently with the existing shower framework would be required, which one can expect to be a highly non-trivial task. 

In terms of the higher-order ingredients  that may be necessary for extending shower accuracy, the criteria set out in Ref.~\cite{Dasgupta:2020fwr} are useful to examine. In particular while NLL accurate configurations involve strong ordering between all pairs of emissions in at least one of two logarithmic variables (e.g. energy and angle), NNLL accuracy requires us to allow for a pair emissions that have comparable values for both logarithmic variables. Meeting this requirement entails the inclusion of {\it{higher-order splitting kernels}} that emerge in the double-soft and triple-collinear kinematic limits \cite{Dokshitzer:1992ip,Campbell:1997hg,
Catani:1998nv,Catani:1999ss}. The inclusion of higher-order kernels in showers has been actively investigated in recent literature, both in the double-soft and triple-collinear limits \cite{Li:2016yez,Hoche:2017hno,Hoche:2017iem,Dulat:2018vuy} (see also related work in Refs.~\cite{Skrzypek:2011zw,Jadach:2016zgk}). In the same context one should perhaps note that the inclusion of higher-order ingredients does not automatically improve the logarithmic accuracy of showers, which depends on the existing shower structure already being NLL accurate and its interplay with the higher-order ingredients. Given this, an investigation of the logarithmic accuracy of a shower following the inclusion of higher-order ingredients should be carried out, along similar lines to the detailed numerical tests presented in Ref.~\cite{Dasgupta:2020fwr}, before NNLL accuracy can reasonably be claimed.

In the current paper we shall carry out analytical studies, relevant to the derivation of specific higher-order shower ingredients, focussing purely on the triple-collinear limit. While the inclusion of the full triple-collinear splitting kernels \cite{Campbell:1997hg,Catani:1998nv} is eventually needed for general NNLL accuracy as mentioned above, we shall initially consider observables such as global event shapes, which are not directly sensitive to the splitting kernels themselves but receive an NNLL collinear contribution from an integral over the splitting kernels.\footnote{This is analogous to the fact that to reach NLL accuracy for global event shapes, one needs to include the CMW coefficient $K$ (also known as $A_2$ in the resummation literature) which is given by an integral of the double-soft correlated emission matrix element, but one does not need the full double-soft matrix element itself.} The coefficient that emerges from the integral of the triple-collinear splitting kernels is known in the resummation literature as $B_2$ \cite{Collins:1981va,Collins:1985xx,Kodaira:1981nh,Kodaira:1982az}. This coefficient governs the intensity of hard-collinear radiation from an initial parton at order $\alpha_s^2$. Its effective inclusion along with that of the resummation coefficient $A_3$ relevant to the soft limit \cite{Moch:2004pa,Vogt:2004mw}, will be a key component of any future NNLL showers.

In this article we seek to better understand the connection between the  $B_2$ coefficient and the triple-collinear splitting kernels by aiming to derive  $B^q_2$, the result for a quark initiated jet, directly from the splitting kernels. In Ref.~\cite{Anderle:2020mxj} we have recently derived the NNLL terms for the groomed jet mass distribution using the modified Mass Drop Tagger (mMDT or equivalently Soft Drop ($\beta=0$)\cite{Dasgupta:2013ihk,Larkoski:2014wba}), directly from the triple-collinear splitting functions. This work included reconstructing $B^q_2$ as well as involving derivation of relevant NNLL jet clustering corrections.\footnote{For related past work from other authors, involving triple-collinear splittings for the case of initial state splittings, we refer the reader to Ref.~\cite{deFlorian:2001zd}.} In the current work we shall aim to extend and generalise the insight that emerged from the study of Ref.~\cite{Anderle:2020mxj}. In particular, in order to facilitate eventual inclusion in parton showers, we shall derive here a version of $B^q_2$ that is differential in the kinematics of a given emission and allows us to project  to a definite phase-space point for a $1 \to 2$ branching. This in turn potentially allows us to include $B^q_2$ by effectively giving an NLO weight to emissions that is correct in the hard-collinear limit and reproduces the standard $B^q_2$, familiar from resummation, upon integration over emission kinematic variables. As we shall demonstrate, examining higher-order collinear branchings in this manner also naturally produces a picture that points to an extension of the CMW coupling scale and scheme beyond the soft limit. Such an extension of the physical coupling ought to be of value for parton shower development but is also, we believe, of intrinsic theoretical interest.

This paper is organised as follows: We start in section \ref{sec:prelim} with a brief reminder of the resummation coefficients, including $B^q_2$ as defined and used in the literature. We also provide a reminder of the triple-collinear splitting functions and the phase-space we use for our calculations, as well as discussing the differential distributions we shall study. Section \ref{sec:results} is dedicated to our results. We start by briefly presenting leading-order results for the distributions in jet mass $\rho$ or equivalently the angle $\theta_g^2$, and energy fraction $z$ for a given emission. In subsections \ref{sec:nfpiece}, \ref{sec:id} and \ref{sec:cfcapure} we describe our order $\alpha_s^2$ results for the $C_F T_R n_f$, $C_F \left(C_F-C_A/2\right)$ and pure $C_F C_A$ terms respectively, all of which arise from the decay of an initially emitted massive parent gluon. The results we obtain are for the differential distributions in the invariant mass of the three parton system $\rho$ and the energy fraction $z$ of the parent gluon emission, as well as results for the distribution in $\theta_g^2$ and $z$, with $\theta_g^2$ being the angle of the parent gluon with respect to the final state quark. In subsection \ref{sec:b2} we use our results to extract the contribution corresponding to $\mathcal{B}^q_2(z)$, i.e. a differential version of $B^q_2$, for each colour channel arising from the gluon decay subprocess. We discuss the structure of the results, including the relationship between  the results for the $\rho$ and $\theta_g^2$ distributions and explain how they are connected. We also obtain a direct connection between the $z$ dependent functions we obtain from our differential distributions, and the NLO non-singlet timelike DGLAP splitting functions. In subsection \ref{sec:web} we combine our results with leading-order (order $\alpha_s$) results, show how our results point towards an extension of the CMW coupling scale and scheme, and may be viewed as an effective extension of the concept of a {\it{web}} \cite{Gatheral:1983cz,Frenkel:1984pz,Laenen:2000ij,Dixon:2008gr} beyond the soft limit. In subsection \ref{sec:ab} we demonstrate how we may extract $\mathcal{B}^q_2(z)$ in the abelian $C_F^2$ channel by noting that it ought to arise from the difference between a calculation performed with the triple-collinear splitting functions (including the one-loop corrections to a $1\to 2$ splitting) and that performed using simply an iteration of successive $1\to 2$ splittings. Finally in section \ref{sec:concl} we draw our conclusions, and comment on prospects for future work. All other technical details relevant to our calculations can  be found in the appendices. 

\section{$B_2^q$, triple-collinear splitting functions and observable definitions}
\label{sec:prelim}
We begin with a brief reminder of the resummation literature in order to introduce the $B_2^q$ coefficient, whose differential version we seek to compute in this article. This is the coefficient that controls the intensity of collinear radiation from a quark at order $\alpha_s^2$, and hence is directly related to the quark Sudakov form factor.  To exemplify this we report below the expression for the Sudakov form factor for the case of transverse momentum related resummation: 
\begin{align}
\label{eq:qtsudakov}
	S(Q,b) = \exp\left(-\int_{\bar{b}^2/b^2}^{Q^2} \frac{d q^2}{q^2} \left[A\left(\alpha_s\left(q^2\right)\right) \ln\frac{Q^2}{q^2} + B\left(\alpha_s\left(q^2\right)\right)\right]\right),\quad \bar{b} = 2e^{-\gamma_E} \ \ ,
\end{align} 
where $b$ is the impact parameter which is the Fourier conjugate of the transverse momentum or related variable. The $A$ function encompasses soft emission effects, while the $B$ function captures the effects of hard-collinear radiation.
Each function admits a perturbative series
\begin{align}
	A(\alpha_s) = \sum_{n=1}^\infty \left(\frac{\alpha_s}{2\pi}\right)^n A_n \ \ ,\quad B(\alpha_s) = \sum_{n=1}^\infty \left(\frac{\alpha_s}{2\pi}\right)^n B_n \ \ .
\end{align}
The perturbative coefficients are then determined by matching the resummation formula onto a fixed-order computation for a given observable \cite{deFlorian:2001zd,deFlorian:2004mp,Davies:1984hs}. The form of $B^q_2$ has long been known to have the following structure \cite{Catani:2000vq,deFlorian:2001zd,deFlorian:2004mp,Banfi:2018mcq}

\begin{align}
\label{eq:defb2}
	B_2^q = - \gamma_q^{(2)} + C_F b_0 X_v , \quad b_0 = \frac{11}{6} C_A - \frac23 T_R n_f \ \ ,
\end{align}
where $b_0$ is the first coefficient of the QCD beta function and $X_v$ is a process and observable dependent constant. Finally, $\gamma_q^{(2)}$ is the end-point contribution of the non-singlet next-to-leading order DGLAP kernel \cite{Ellis:1996mzs}
\begin{align}
\label{eq:defgammaq}
	\gamma_q^{(2)} = C_F^2 \left(\frac38 - \frac{\pi^2}{2} + 6 \zeta(3) \right) + C_F C_A \left(\frac{17}{24} + \frac{11 \pi^2}{18} - 3 \zeta(3)\right) - C_F T_R n_f \left(\frac16 + \frac{2\pi^2}{9}\right) \ \ .
\end{align}

Having provided a reminder of the $B_2^q$ coefficient, we turn to the triple-collinear splitting functions that we shall use throughout this paper. The polarisation-averaged triple-collinear splitting functions we consider here were first derived in Refs.~\cite{Campbell:1997hg,Catani:1998nv}. For an initial quark there are four distinct splitting functions to consider. In the notation of Ref.~\cite{Catani:1998nv}, these are: 
\begin{itemize}
	\item $\langle \hat{P}_{\bar{q}^\prime_1 q^\prime_2 q_3} \rangle$ corresponding to a $1 \to 3$ quark splitting involving non-identical quark flavours, associated with a $C_F T_R$ colour factor.
	\item $\langle \hat{P}^{\text{(id)}}_{\bar{q}_1 q_2 q_3} \rangle$  representing an additional interference contribution  coming from identical quark flavours, with a $C_F \left(C_F-C_A/2 \right)$ colour factor.
	\item  $\langle \hat{P}^{\text{(nab)}}_{g_1 g_2  q_3} \rangle$ the non-abelian contribution leading to two final-state gluons and a quark, with a $C_F C_A$ colour factor.
	\item  $\langle\hat{P}^{\text{(ab)}}_{g_1 g_2 q_3} \rangle$ the abelian contribution also involving two final-state gluons and a quark but arising from independent gluon emission off a quark, with a $C_F^2$ colour factor.
\end{itemize}

We shall work in terms of the energy fractions of the three collinear partons $z_i$, satisfying $\sum_i z_i =1$, and shall label the angles between partons $i$ and $j$ as $\theta_{ij}$, such that in the collinear approximation, relevant to this work, $\theta_{ij}  \ll 1$. The  splitting functions are reported in Ref.~\cite{Catani:1998nv} as functions of $z_i$ and invariants $s_{ij}$ and $s_{123}$ where $s_{ij} = (p_i+p_j)^2 \approx E^2 z_i z_j \theta_{ij}^2$ where $E$ is the energy of the initial quark and $s_{123}=(p_1+p_2+p_3)^2 = s_{12}+s_{13}+s_{23}.$
The  triple-collinear phase-space in $d=4-2\epsilon$ dimensions may be expressed in the form \cite{Gehrmann-DeRidder:1997fom}

\begin{equation}
\label{eq:tcphasesp}
\text{d} \Phi_3 = \frac{1}{\pi}\left(E\right)^{4-4\epsilon} \frac{1}{\left(4\pi\right)^{4-2\epsilon} \Gamma\left(1-2\epsilon\right)}
 \text{d}z_2 \text{d}z_3  \text{d} \theta_{13}^2    \text{d} \theta_{23}^2
\text{d} \theta_{12}^2\, (z_1 z_2 z_3)^{1-2\epsilon}\, \Delta^{-1/2-\epsilon}  \, \theta (\Delta),
\end{equation}
where the Gram determinant $\Delta$ is defined as
\begin{equation}
\Delta = 4 \theta_{13}^2 \theta_{23}^2 - \left(\theta_{12}^2-\theta_{23}^2-\theta_{13}^2\right)^2.
\end{equation}

Additionally, in Ref.~\cite{Anderle:2020mxj} we used a set of variables, which can be referred to as ``web variables'', to parametrize the triple-collinear phase space. 
The web variables will again prove essential to obtain analytic expressions for the distributions we are interested in. In appendix~\ref{app:web} we recapitulate the phase space and recollect the physical meaning of the different variables.

\begin{figure}[h]
	\centering
	\begin{tikzpicture}[scale=2.5] 
		
		\coordinate (bq1) at (0,0);
		\coordinate (bq2) at (1,0); 
		\coordinate (bq3) at (2.5,0);
		
		\coordinate (bg1) at (1.2,0);
		\coordinate (eg1) at (1.5,0.6);
		\coordinate (bg2) at (1.5,0.6);
		\coordinate (eg2) at (1.3,1.0);
		\coordinate (bg3) at (1.5,0.6);
		\coordinate (eg3) at (1.9,0.8);
		
		\draw [quark] (bq1) -- (bq2);
		\draw [gluon] (bg1) -- (eg1) ;
		\node at (1.4,0.1) {\small $\theta_g$};
		\draw [quark] (bg2) -- (eg2) node [pos=1,left] {\small $z_1=(1-z)z_p$};
		\node at (1.58,0.8) {\small $\theta_{12}$};
		\draw [quark] (eg3) -- (bg3);
		 \node at (2.55,0.8) {\small $z_2=(1-z)(1-z_p)$};
		\draw [quark] (bq2) -- (bq3) node
		[pos=1,right] {\small $z_3 = z$} ;

	\end{tikzpicture}
	\caption{The Feynman diagram representing gluon decay to a $q\bar{q}$ pair, where the quark from the gluon decay is either identical or non-identical to the initiating quark. The mapping of the momentum fractions to a set of independent variables is explained in the text.}\label{fig:triple-collinear-fermion}
	\label{fig:qqqbar}
\end{figure}
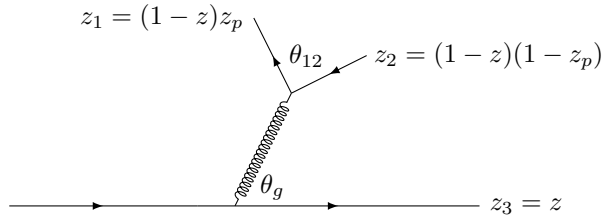

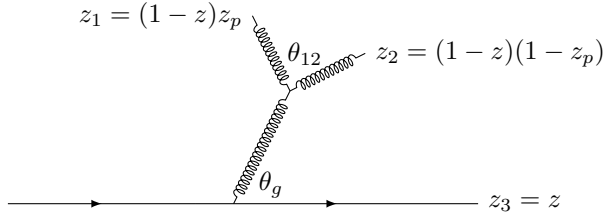
\begin{figure}[h]
	\centering
	\begin{tikzpicture}[scale=2.5] 
		
		\coordinate (bq1) at (0,0);
		\coordinate (bq2) at (1,0); 
		\coordinate (bq3) at (2.5,0);
		
		\coordinate (bg1) at (1.2,0);
		\coordinate (eg1) at (1.5,0.6);
		\coordinate (bg2) at (1.5,0.6);
		\coordinate (eg2) at (1.3,1.0);
		\coordinate (bg3) at (1.5,0.6);
		\coordinate (eg3) at (1.9,0.8);
		
		\draw [quark] (bq1) -- (bq2);
		\draw [gluon] (bg1) -- (eg1) ;
		\node at (1.4,0.1) {\small $\theta_g$};
		\draw [gluon] (bg2) -- (eg2) node [pos=1,left] {\small $z_1=(1-z)z_p$};
		\node at (1.58,0.8) {\small $\theta_{12}$};
		\draw [gluon] (bg3) -- (eg3) node [pos=1,right] {\small $z_2=(1-z)(1-z_p)$};
		\draw [quark] (bq2) -- (bq3) node
		[pos=1,right] {\small $z_3 = z$} ;

	\end{tikzpicture}
	\caption{The Feynman diagram representing the gluon decay $C_F C_A$ channel. The mapping of the momentum fractions to a set of independent variables is explained in the text. The angle $\theta_g$ is that between the parent gluon and the final-state quark.}\label{fig:triple-collinear-cfca}
	\label{fig:qggnab}
\end{figure}

\begin{figure}[h]
	\centering
	\begin{tikzpicture}[scale=2.5] 
		
		\coordinate (bq1) at (0,0);
		\coordinate (bq2) at (1,0); 
		\coordinate (bq3) at (2.5,0);

		\coordinate (bg1) at (.8,0);
		\coordinate (eg1) at (1.2,0.6);
		\coordinate (bg2) at (1.65,0);
		\coordinate (eg2) at (2.05,0.6);
		
		\draw [quark] (bq1) -- (bq2);
		\draw [gluon] (bg1) -- (eg1) node [pos=1,right] {\small $z_1=1-z$} ;
		\node at (1.18,0.1) {\small $\theta_{1,2+3}$};
		\draw [gluon] (bg2) -- (eg2) node [pos=1,right] {\small $z_2=z(1-z_p)$};
		\node at (1.95,0.1) {\small $\theta_{23}$};
		\draw [quark] (bq2) -- (bq3) node
		[pos=1,right] {\small $z_3 = z z_p$};

	\end{tikzpicture}
\caption{The Feynman diagram representing the gluon emission $C_F^2$ channel. The mapping of the momentum fractions to a set of independent variables is explained in the text.}\label{fig:triple-collinear-cf2}
\label{fig:qggab}
\end{figure}
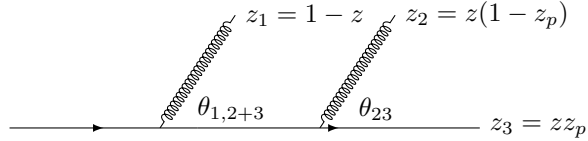

In Figures \ref{fig:qqqbar} -- \ref{fig:qggab} we illustrate the splitting sub-processes involved here, namely a gluon decay contribution showing a $q \to qg$ splitting followed by gluon splitting to a $q\bar{q}$  pair (Figure \ref{fig:qqqbar}) and a $gg$ pair (Figure \ref{fig:qggnab}) as well as the abelian contribution to $q\to qgg$ with independent gluon emission off a quark (Figure \ref{fig:qggab}). The splitting process shown in Figure \ref{fig:qqqbar} gives both the identical and non-identical fermion splitting functions. Figures \ref{fig:qqqbar} -- \ref{fig:qggab} also illustrate our change of independent variables from $z_2, z_3$ to $z$ and $z_p$  which are the variables associated to successive splittings, and are defined so  that in the limit of strong angular-ordering between successive branchings the triple-collinear splitting functions reduce (after azimuthal integration for the gluon decay channels) to a product of leading-order splitting functions  in $z$ and $z_p$ respectively. 

Finally we discuss  the quantities we study here.  These include the double differential distribution in $\rho =s_{123}/E^2$ and $z$, where $\rho$ is  the normalised invariant mass of the three parton system that arises from the triple-collinear splitting of a quark jet with energy $E$, and the splitting variable $z$ may be associated to an initial splitting as illustrated in Figures  \ref{fig:qqqbar} -- \ref{fig:qggab}.\footnote{This initial splitting will also set the small angular scale which defines the collinearity of the problem.} For the gluon decay contributions, as should be evident from Figures  \ref{fig:qqqbar} and \ref{fig:qggnab}, the variable $z$ also corresponds to the energy fraction of the final-state quark so that  $1-z$ represents the energy fraction associated to the ``parent'' gluon.  In addition to the $\rho$ distribution we shall also study the distribution differential in $z$ and angle $\theta_g$ of the parent gluon. A comparison of the two distributions shall give further insight into the general structure of the result. For the abelian gluon emission process in Figure \ref{fig:qggab}, we shall fix $\theta_{13}=\theta  \ll 1$, the angle of emission $1$ wrt the {\it{final}} quark, which shall set the collinearity, and then study the NLO structure induced by a smaller angle emission labelled 2,  with angle $\theta_{23}  <  \theta$.

We integrate the splitting functions over phase-space in $d=4-2\epsilon$ dimensions to obtain real emission contributions that contain poles in  $\epsilon$ which reflect singularities that cancel when we combine with virtual corrections. The integrals we carry out are generically of the form
\begin{equation}
\frac{v}{\sigma_0}\frac{d\sigma}{dv} = \int \text{d}\Phi_3(z_i,\theta_{ij}) \frac{\left(8\pi \alpha \mu^{2\epsilon}\right)^2}{s^2_{123}} \langle \hat{P} \rangle \, v\, \delta \left ( v -v \left(z_i,\theta_{ij} \right)\right)
\end{equation}
where  $v$ denotes the quantity we hold fixed, $ \langle \hat{P} \rangle$ denotes the different $1\to3$ splitting functions mentioned above, $\alpha$ denotes the bare QCD coupling, and $\sigma_0$ is the Born cross section. Our results will be expressed in terms of a renormalised $\overline{\text{MS}}$ coupling $\alpha_s$, given by the relation
\begin{equation}
\label{eq:renorm}
\mu^{2\epsilon} \alpha = S^{-1}_\epsilon\mu_R^{2\epsilon} \alpha_s(\mu_R^2) +\mathcal{O}(\alpha_s^2),
\end{equation}
where  we have the standard $\overline{\mathrm{MS}}$ factor
\begin{equation}
\label{eq:msbar}
S_\epsilon = (4\pi)^\epsilon e^{-\epsilon \gamma_E},
\end{equation}
and we choose $\mu_R =E$ (the energy of the hard initiating parton).  For the results that follow we define $\alpha_s  \equiv \alpha_s (E^2)$ in the  $\overline{\text{MS}}$ scheme.

\section{Results}
\label{sec:results}
We note that the leading-order collinear limit result for the $\rho$ distribution is given by considering a single collinear gluon emission from a quark, with energy fraction $z$ and angle $\theta_g$ so that $\rho = z(1-z)\theta_g^2$. This result is equivalent to that for fixed $\theta_g$ and $z$ so that we have
\begin{equation}
\label{eq:lo}
\frac{\rho}{\sigma_0} \frac{d^2\sigma^{(1)}}{d\rho\, dz} = \frac{\theta_g^2}{\sigma_0}  \frac{d^2\sigma^{(1)}}{d\theta_g^2 \, dz} = \frac{C_F \alpha_s}{2\pi} \left(\frac{1+z^2}{1-z}\right) \ \ .
\end{equation}
\subsection{$C_F T_R  n_f$ terms}
\label{sec:nfpiece}
At order $\alpha_s^2$ we  start by  considering  $C_F T_R n_f$ terms both for fixed invariant-mass of the three-parton system $\rho$ and $z$, as well as for fixed $\theta_g$ and $z$ where $\theta_g$ is the angle of the parent gluon with respect to the final quark. For the former case we have already obtained a previous fully analytic result as part of our study of the modified Mass Drop Tagger's jet mass distribution. This was reported in Ref.~\cite{Anderle:2020mxj} and we shall analyse its structure in more detail here, together with that of the $\theta_g^2$ distribution calculated here.

\subsubsection{Fixed invariant-mass}
The order $\alpha_s^2$ result for the  $\rho$ distribution, in the collinear limit, was calculated in Ref.~\cite{Anderle:2020mxj}. Two separate calculations were performed there. 
Firstly we carried out a part analytic and part numerical computation where the divergent terms were computed analytically and a finite remainder computed through numerical integration in four dimensions. Secondly we computed a fully analytical result using a parametrisation of the phase space based on web variables. The two calculations were found to be in perfect agreement and we report below the fully analytical result \cite{Anderle:2020mxj}:

\begin{align}\label{eq:rhocftf}
\left(\frac{\rho}{\sigma_0} \frac{d^2\sigma^{(2)}}{d\rho\, dz}\right)^{C_F T_R n_f} = C_F T_R n_f  \left(\frac{\alpha_s}{2\pi} \right)^2 \left( \frac{1+z^2}{1-z} \left( \frac{2}{3}   \ln \left(\rho(1-z)\right)   -\frac{10}{9} \right)-\frac{2}{3}(1-z)\right) \ \ .
\end{align}

\subsubsection{Fixed parent angle}
We now study the distribution in the angle of the parent gluon and $z$. The angle of the parent can be straightforwardly expressed in terms of our phase-space variables and is given, in a collinear approximation, by 
$\theta_g^2 \simeq z_p \theta_{13}^2+(1-z_p) \theta_{23}^2-z_p(1-z_p)\theta_{12}^2$ with variables as illustrated in Figure \ref{fig:qqqbar}.  We note that in the limits $z_p \to 0$ or $z_p \to 1$, where all the energy is carried by a given offspring parton, $\theta_g^2$ reduces to the angle of the energetic offspring parton ($1$ or $2$) wrt parton $3$ i.e. the quark. In the collinear limit $\theta_{12} \to 0, \theta_{13} \to \theta_{23}$, $\theta_g^2$ reduces to the angle between the direction of either collinear parton and the quark. 

We follow the same strategy as the one adopted in Ref.~\cite{Anderle:2020mxj} for the $\rho$ distribution and perform the calculation using two different methods: firstly a partly analytical calculation with a finite remainder evaluated numerically and secondly a fully analytical calculation using web variables. 

In the partly analytic approach, integrating the triple-collinear splitting function $\langle \hat{P}_{\bar{q}_1^\prime q^\prime_2 q_3} \rangle$ at fixed $z$ and $\theta_g$ we obtain a result that has only a $\frac{1}{\epsilon}$ pole which originates in the collinear divergence as the angle between gluon offspring partons  $\theta_{12} \to 0$. The collinear pole is multiplied as usual by an $\epsilon$ dependent factor and on performing an $\epsilon$ expansion we obtain a pure $\frac{1}{\epsilon}$ pole term and associated finite terms.

Further, on subtracting the leading collinear divergent piece from the full integrand, we also obtain a finite term in addition to the collinear divergent term, that can be computed by integration in four dimensions, i.e. setting $\epsilon \to 0$. This finite term has two components: a term that behaves  as $1/(1-z)$ i.e. is singular in the limit of a soft parent and we can extract analytically, and a term that is regular as $z\to1$ which we can leave to numerical integration.

In this approach, the analytically determined component of our answer, which contains the collinear divergence, can be expressed as 

\begin{equation}
\left(\frac{\theta_g^2}{\sigma_0} \frac{d^2\sigma^{(2)}}{d\theta_g^2 dz}\right)^{C_F T_R n_f, \text{div.}} = C_F T_R n_f  \left(\frac{\alpha_s}{2\pi} \right)^2 \left( \frac{1+z^2}{1-z}\left(- \frac{2}{3\epsilon}-\frac{10}{9}+\frac{4}{3}  \ln (z(1-z) ^2 \theta_g^2)\right)-\frac{7}{9}(1-z)\right).
\end{equation}

The additional finite term that can be evaluated in four dimensions turns out to be identical to the corresponding result for the $\rho$ distribution.  This term takes the form below, confirmed by both our numerical evaluation and our separate fully analytic evaluation (see appendix~\ref{app:cftftheta})
\begin{equation}
\left(\frac{\theta_g^2}{\sigma_0} \frac{d^2\sigma^{(2)}}{d\theta_g^2 dz}\right)^{C_F T_R n_f, \text{fin.}}  = C_F T_R n_f  \left(\frac{\alpha_s}{2\pi} \right)^2 \left( \frac{2}{3} \, \frac{1+z^2}{1-z} \ln z+  \frac{7}{9}(1-z)\right) \ \ .
\end{equation}

Finally we need the one-loop virtual correction to the $1 \to 2$ $q \to qg$ splitting which reads:
\begin{equation}\label{eq:cftfvirt}
\left(\frac{\theta_g^2}{\sigma_0} \frac{d^2\sigma}{d\theta_g^2 dz}\right)^{C_F T_R n_f, \text{virt.}} = C_F T_R n_f  \left(\frac{\alpha_s}{2\pi} \right)^2 \left(\frac{2}{3} \frac{1+z^2}{1-z} \left(\frac{1}{\epsilon}-\ln(z^2 (1-z)^2  \theta_g^2) \right)  -\frac{2}{3}(1-z)\right) \ \ .
\end{equation}

Combining all terms we get
\begin{equation}
\label{eq:thetadist}
\left(\frac{\theta_g^2}{\sigma_0} \frac{d^2\sigma^{(2)}}{d\theta_g^2\, dz}\right)^{C_F T_R n_f} =  C_F T_R n_f  \left(\frac{\alpha_s}{2\pi} \right)^2 \left( \frac{1+z^2}{1-z} \left( \frac{2}{3}   \ln \left(z(1-z)^2\theta_g^2 \right)   -\frac{10}{9} \right)-\frac{2}{3}(1-z)\right) \ \ .
\end{equation}

A notable feature of our result is that it can be obtained from that for the $\rho$ distribution via the replacement $\rho \to z(1-z) \theta_g^2$ which is an exact relationship between the two observables at leading order i.e. for the emission of an on-shell gluon. The reason for this is simply the fact that the observable dependence of our result comes solely from the  divergent limit when $\theta_{12} \to 0$,  where IRC safe observables tend to their leading-order form. The finite terms that are computable in four dimensions are instead observable independent. More specifically, for some fixed value of observable $v$, once divergent contributions are removed we are left with genuine triple-collinear configurations with three energetic partons with comparable opening angles of order $\theta^2 \sim v$. The overall $1/\theta^2$ scaling of the  triple-collinear splitting functions then implies that $v d\sigma/\sigma_0 dv$ is independent of $v$.

\subsection{$C_F \left(C_F-\frac{C_A}{2} \right)$ terms}
\label{sec:id}
Next we examine the colour suppressed identical fermion term also arising from gluon splitting to $q\bar{q}$. This is a purely finite term and can be calculated in four dimensions. In Ref.~\cite{Anderle:2020mxj} we numerically computed the contribution of this term to $\rho \frac{d\sigma}{d\rho}$ as part of our calculations for the mMDT jet mass. However,  in Ref.~\cite{Anderle:2020mxj}, we did not  determine the  $z$ dependence of our result as we only required the integral over $z$. After the computational steps explained in appendix~\ref{app:id}, we provide a  fully analytical result for the $\rho$ distribution also differential in $z$. We also carry out a numerical calculation for fixed $\theta_g^2$ where we continue to label as $\theta_g$ the angle $\theta_{1+2,3}$, the angle between the direction of the $1,2$ parton pair and the quark labelled 3. However we note that it is no longer possible to interpret this angle unambiguously as the angle of the parent gluon due to the identical quarks present in the final state.  As expected, for pure finite terms, the results for the $\rho$ and $\theta_g^2$ distributions coincide so that we obtain 
\begin{align}\label{eq:ident}
\left(\frac{\rho}{\sigma_0} \frac{d^2\sigma^{(2)}}{d\rho\, dz}\right)^{\text{(id.)}} &= \left(\frac{\theta_g^2}{\sigma_0} \frac{d^2\sigma^{(2)}}{d\theta_g^2\, dz}\right)^{\text{(id.)}}= C_F \left(C_F-\frac{C_A}{2}\right)  \left( \frac{\alpha_s}{2\pi}\right)^2 \mathcal{P}^{\text{(id.)}}(z),\\ \nonumber
\mathcal{P}^{\text{(id.)}}(z) &= \left(4z-\frac{7}{2} \right)+\frac{5z^2-2}{2(1-z)} \ln z+\frac{1+z^2}{1-z}\left(\frac{\pi^2}{6} -\ln z \ln (1-z)-\text{Li}_2(z)\right).
\end{align}
In deriving $\mathcal{P}^{\text{(id.)}}(z)$ we have defined $z_3 =z$ as usual but the same result is obtained for the distribution in $z_2$, as one may readily anticipate for identical particles. 

\subsection{Pure $C_F C_A$ term}
\label{sec:cfcapure}
In Ref.~\cite{Anderle:2020mxj} we provided a calculation for the $C_F C_A$ contribution for the $\rho$ distribution, that arises from the decay to $gg$ of a parent gluon emitted off a quark. Our results were obtained in a form which was partly analytic, while finite terms computable in four dimensions were obtained through numerical integration. Here we report a fully analytic result with the details of the calculation left to appendix~\ref{app:cfcatheta}. 
We obtain
\begin{align}\label{eq:rhodistcfca}
\left(\frac{\rho}{\sigma_0} \frac{d^2\sigma^{(2)}}{d\rho\, dz}\right)^{\text{nab.}} = C_F C_A \left(\frac{\alpha_s}{2\pi}\right)^2 \mathcal{P}^{(\text{nab.})}(z;\rho) \ \ ,
\end{align}
where
\begin{align}
\label{eq:rhodistcfca2}
	\nonumber
	\mathcal{P}^{(\text{nab.})} (z;\rho) = \left(\frac{1+z^2}{1-z} \right)&\left(-\frac{11}{6} \ln \left(\rho(1-z)\right) +\frac{67}{18}-\frac{\pi^2}{6} +\ln^2z+ \mathrm{Li}_2\left(\frac{z-1}{z}\right)+2\, \mathrm{Li}_2(1-z)\right)+\\ 
	&+\frac{3}{2} \frac{z^2 \ln z}{1-z} +\frac{1}{6}(8-5z) \ \ .
\end{align}
The results for the  $\theta_g^2$ distribution give a corresponding function $\mathcal{P}^{\text{nab}.} (z;\theta_g^2) $. As observed for the $C_F T_R n_f$ term the observable dependence in $\rho/\sigma_0\, d\sigma/d\rho$ arises from the collinear divergent limit so that one may obtain  the $\theta_g^2$ result via a simple substitution for $\rho$. Hence we have
\begin{equation}\label{eq:thetadistcfca}
\left(\frac{\theta_g^2}{\sigma_0} \frac{d^2\sigma^{(2)}}{d\theta_g^2 \, dz}\right)^{\text{nab.}} = C_F C_A \left(\frac{\alpha_s}{2\pi}\right)^2 \mathcal{P}^{(\text{nab.})}(z;\theta_g^2) \ \ ,
\end{equation}
\begin{equation}
\label{eq:thetadistcfca2}
	\mathcal{P}^{(\text{nab.})} (z; \theta_g^2) = \mathcal{P}^{(\text{nab.})} (z; \rho=z(1-z)\theta_g^2) \ \ .
\end{equation}

\subsection{$\mathcal{B}_2^{q}(z)$ and comments on general structure}
\label{sec:b2}
We now study the link between our results and $B_2^{q}$, the parameter in the quark form factor that relates to the intensity of collinear radiation from a quark at $\mathcal{O}(\alpha_s^2)$. We discuss also  the connection between our results and the NLO timelike splitting functions \cite{Ellis:1996mzs}. Finally we show how to combine our results with the corresponding leading-order results, recovering the expected behaviour in the soft limit and giving a simple picture for $z$ dependent corrections beyond the soft limit.

\subsubsection{Extracting $B^q_2$}
Let us start with the $\theta_g^2$ distribution, whose $C_F T_R n_f$ term is reported in Eq.~\eqref{eq:thetadist}. In the soft limit, $z \to 1$ the result reduces to a familiar one:
\begin{align}
\label{eq:thetadistsoft}
\left(\frac{\theta_g^2}{\sigma_0} \frac{d^2\sigma^{(2)}}{d\theta_g^2\, dz}\right)^{\text{soft},C_F T_R n_f}  &=C_F T_R n_f  \left(\frac{\alpha_s}{2\pi} \right)^2  \frac{2}{1-z} \left( \frac{2}{3}   \ln \left((1-z)^2\theta_g^2 \right)   -\frac{10}{9} \right) \\ 
&= C_F \frac{2}{1-z}  \left(\frac{\alpha_s}{2\pi} \right)^2 \left(-b_0^{(n_f)} \ln \frac{k_t^2}{E^2} +K^{(n_f)} \right) \ \ ,
\label{eq:thetadistsoft2}
\end{align}
where we note the presence of a soft divergence as $z \to 1$, reflecting the singular behaviour of the $q \to qg$ splitting function. In writing the second line we have introduced the transverse momentum of the emitted parent gluon $k_t^2 =E^2 (1-z)^2 \theta_g^2$ wrt the final quark direction. We have also defined $b_0^{(n_f)}$, which is the $n_f$ part of the first perturbative coefficient of the QCD beta function (see Eq.~\eqref{eq:defb2}) and $K^{(n_f)}$ is the $n_f$ term in the CMW constant \cite{Catani:1990rr}
   \begin{align}
   	K = \left(\frac{67}{18}-\frac{\pi^2}{6} \right)C_A -\frac{10}{9} T_R n_f  \ \ .
   \end{align}
   The terms appearing in Eq.~\eqref{eq:thetadistsoft} are essentially related to NLL rather than NNLL structure, leading to the well known soft-limit prescriptions for the scale and physical scheme for the strong coupling. 
   Indeed in terms of the type of decomposition into soft and collinear pieces presented in Eq.~\eqref{eq:qtsudakov} these terms are all associated to the $``A"$ series of coefficients owing to the presence of the soft divergence. Here we wish to obtain  the pure collinear NNLL structure and hence we shall subtract off these terms.  We shall also remove the remaining piece of  the $\ln \theta_g^2$ term $\propto-(1+z) \ln \theta_g^2$ as this term is also an NLL contribution. Indeed, the $-(1+z) \ln\theta_g^2$ yields after $z$ integration $-3/2 \ln \theta_g^2$ which makes clear its association with the $B_1$ hard-collinear coefficient, rather than $B_2$.\footnote{Specifically we can see from Eq.~\eqref{eq:qtsudakov} the presence of a term $B_1 \alpha_s(q^2)$ which, with the transverse momentum $q^2 \propto \theta_g^2$, results in the presence of a term of the form $3/2 \,\alpha_s^2 \, b_0 \ln \theta_g^2$ as seen in our results after $z$ integration.}   
 
 Hence we obtain
 \begin{align}
   \label{eq:Bzdef}
\mathcal{B}^{q,n_f}_2 (z;\theta_g^2)& =\left(\frac{\theta_g^2}{\sigma_0} \frac{d^2\sigma^{(2)}}{d\theta_g^2\, dz}\right)^{C_F T_R n_f} -\left  [\left(\frac{\theta_g^2}{\sigma_0} \frac{d^2\sigma^{(2)}}{d\theta_g^2\, dz}\right)^{\text{soft},C_F T_R n_f} -C_F T_R n_f  \frac{2}{3} \left(\frac{\alpha_s}{2\pi} \right)^2 (1+z) \ln \theta_g^2 \right]\\
&= C_F T_R n_f  \left(\frac{\alpha_s}{2\pi} \right)^2  \left( \frac{1+z^2}{1-z} \, \frac{2}{3} \ln z -(1+z) \, \left(\frac{2}{3} \ln (1-z)^2-\frac{10}{9}\right)-\frac{2}{3}(1-z) \right),
\end{align}
where the subtracted terms are placed in square brackets and leave an NNLL pure collinear result denoted by $\mathcal{B}^{q,n_f}_2 (z;\theta_g^2)$, which is a differential version of the $n_f$ term of $B^q_2$.

A similar result can of course be written for the $\rho$ distribution by again removing soft enhanced and NLL in $\rho$ terms . Indeed, as we have noted before, the result for the $\rho$ distribution can be reached from that for the $\theta_g^2$ distribution via the one-gluon relation between observables i.e. via the replacement $\theta_g^2   \to \rho/(z(1-z)$. This correspondence ultimately results in the relationship:
\begin{equation}
\mathcal{B}^{q,n_f}_2 (z;\rho) = \mathcal{B}^{q,n_f}_2 (z;\theta_g^2) - C_F T_R n_f \left(\frac{\alpha_s}{2\pi}\right)^2 \left(\frac{1+z^2}{1-z} \, \frac{2}{3} \ln z - (1+z) \frac{2}{3} \ln(1-z)\right) .
\end{equation}

We can now perform the integrals over $z$ to write the results in a standard form:
\begin{align}
\label{eq:B2int}
\nonumber
B^{q,\rho,n_f}_2 &=\left(\frac{2\pi}{\alpha_s}\right)^2\int_0^1  dz\, \mathcal{B}^{q,n_f}_2 (z;\rho) \\
&=  C_F T_R n_f  \, \frac{5}{2} = -\gamma_q^{(2,n_f)}+C_F b^{(n_f)}_0 X_\rho  \ \ ,\\ \nonumber
B^{q,\theta_g^2,n_f}_2 &=\left(\frac{2\pi}{\alpha_s}\right)^2 \int_0^1  dz\, \mathcal{B}^{q,n_f}_2 (z;\theta_g^2)\\
 & = C_F T_R n_f  \left(\frac{9}{2}-\frac{2\pi^2}{9} \right) =  -\gamma_q^{(2,n_f)}+C_F b^{(n_f)}_0 X_{\theta_g^2}  \ \ ,
\label{eq:B2int2}
\end{align}
where $\gamma_q^{(2,n_f)}$ is the $n_f$ part of the endpoint contribution to the DGLAP splitting kernels given in Eq.~(\ref{eq:defgammaq}) and we determine the observable dependent constants
  \begin{align}
   X_\rho = \frac{ \pi^2}{3}-\frac72, \quad X_{\theta_g^2} = \frac{2 \pi^2}{3}-\frac{13}{2} \ \ .
   \end{align} 
 Notice that we inserted a factor of $(2\pi/\alpha_s)^2$ so that our extraction of $B_2$ agrees with the standard definition, i.e. Eq.~\eqref{eq:defb2}.. The above results are in line with the expected form of the  $B_2$ coefficients (see Eq.~\eqref{eq:defb2}) and hence consistent with previous observations in the literature \cite{deFlorian:2004mp} that $B_2$ is always related to the endpoint terms of the DGLAP splitting kernels and additional $b_0  X$ terms where $X$ depends on the observable. Our result in Eq.~\eqref{eq:B2int} for the $\rho$ distribution is in agreement with the collinear NNLL result in the literature for the plain jet mass \cite{Banfi:2018mcq,Chien:2010kc} as we also pointed out in Ref.~\cite{Anderle:2020mxj}. The result for the $\theta_g^2$ case has been computed here for the first time, to our knowledge.

For the identical particle contribution there is no soft contribution to be removed nor any observable dependent $b_0 X$ terms. The result derived in Eq.~\eqref{eq:ident} corresponds to the pure NNLL constant and hence we have 
\begin{equation}
\label{eq:id}
\mathcal{B}^{q,(\text{id.})}_2(z) = C_F \left(C_F-\frac{C_A}{2} \right) \left(\frac{\alpha_s}{2\pi}\right)^2 \mathcal{P}^{\text{(id.)}}(z) \ \ ,
\end{equation}
which gives
\begin{equation}
\label{eq:b2intid}
B^{q,(\text{id.})}_2= \left(\frac{2\pi}{\alpha_s}\right)^2 \int_0^1 dz \, \mathcal{B}^{q,(\text{id.})}_2(z) = C_F \left(C_F-\frac{C_A}{2} \right)  \left( \frac{13}{4}-\frac{\pi^2}{2}+2 \zeta(3) \right) \ \ .
\end{equation}
The above analytical result is consistent with our previous numerical evaluation in the context of the Soft Drop jet mass \cite{Anderle:2020mxj}.\footnote{The result here is for a single emitting leg which results in a factor of 1/2 relative to the calculations of Ref.~\cite{Anderle:2020mxj}.}

Next we turn to the non-abelian contribution to $q \to q gg$, involving the decay of the parent gluon to a pair of gluons, $g \to gg$. An exactly analogous procedure can be followed as for the $g \to q\bar{q}$ decay to remove soft and higher logarithmic order (i.e. NLL)  contributions. In particular, in the soft limit we obtain the result given in Eq.~\eqref{eq:thetadistsoft2} with $b_0^{(n_f)}$ and $K^{(n_f)}$ replaced by the corresponding $C_A$ terms. We thus derive the results below:

\begin{multline}
\label{eq:b2catheta}
\mathcal{B}^{q,(\text{nab.})}_2 (z;\theta_g^2)= C_F C_A \left(\frac{\alpha_s}{2\pi} \right)^2 \left ((1+z) \left(\frac{11}{6} \ln(1-z)^2-\frac{67}{18}+\frac{\pi^2}{6}  \right) +\frac{3}{2} \frac{z^2 \ln z}{1-z}+\frac{8-5z}{6}\right. \\ \left.+\frac{1+z^2}{1-z} \left( -\frac{11}{6} \ln z +\ln^2 z + \mathrm{Li}_2 \left(\frac{z-1}{z} \right)+2\mathrm{Li}_2(1-z) \right) \right),
\end{multline}
and 
\begin{equation}
\label{eq:b2carho}
\mathcal{B}^{q,(\text{nab.})}_2 (z;\rho) = \mathcal{B}^{q,(\text{nab.})}_2 (z;\theta_g^2) +\frac{1+z^2}{1-z} \, \frac{11}{6}\ln z-\frac{11}{6} (1+z) \ln(1-z).
\end{equation}
We note that the sum of dilogarithms multiplying the LO splitting function $p_{qq}(z)$, on the second line of Eq.~\eqref{eq:b2catheta}, is regular as $z \to 1$ and has a small $z$ limit given by $-\frac{1}{2} \ln^2 z+\frac{\pi^2}{6}+\mathcal{O}(z).$

We then have the following integrated results:

\begin{align}
\label{eq:B2intCA}
B^{q,\rho,(\text{nab.})}_2&=\left(\frac{2\pi}{\alpha_s}\right)^2 \int_0^1  dz\, \mathcal{B}^{q,(\text{nab.})}_2 (z;\rho)  =  C_F C_A  \left( -\frac{11}{2}-\frac{\pi^2}{4}+4 \zeta(3)\right)\ \ , \\
B^{q,\theta_g^2,(\text{nab.})}_2 &=\left(\frac{2\pi}{\alpha_s}\right)^2 \int_0^1  dz\, \mathcal{B}^{q,(\text{nab.})}_2 (z;\theta_g^2) = C_F C_A  \left(-11+\frac{13}{36}\pi^2+4\zeta(3)   \right) \ \ .
\label{eq:B2inntCA2}
\end{align}

As we would expect the difference between the results for $\rho$ and $\theta_g^2$ distributions comes only from the $C_F \beta_0 X$ observable dependent terms with values of $X$ already identified in Eq.~\eqref{eq:B2int}.
Thus we have
\begin{equation}
\label{eq:xtheta}
B^{q,\rho,(\text{nab.})}_2-B^{q,\theta_g^2,(\text{nab.})}_2 = C_F C_A  \left( \frac{11}{2}-\frac{11}{18} \pi^2\right) =C_F b_0^{(C_A)} \left(X_\rho-X_{\theta_g^2}\right),
\end{equation}
where $b_0^{(C_A)} =\frac{11}{6} C_A.$  Note however that the integrated results given  in Eq.~\eqref{eq:B2intCA} do not yet fully correspond to the standard form given in Eq.~\eqref{eq:defb2}. In particular in order to fully recover the $C_F C_A$ term of the DGLAP endpoint contribution, $\gamma_q^{(2,C_A)}$, we need to combine the $C_F C_A$ results from Eq.~\eqref{eq:B2intCA} with the identical particle contribution computed in Eq.~\eqref{eq:b2intid} which, by virtue of its $C_F(C_F-C_A/2)$ colour factor, contributes to both $C_F^2$ and $C_F C_A$ colour channels. Defining $B^{q,(\text{id.}),C_A}_2$ as the $C_F C_A$ piece of the identical particle term in Eq.~\eqref{eq:b2intid}, and taking the example of the $\rho$ distribution, we have\footnote{Identical considerations of course hold for the $\theta_g^2$ distribution with $X_\rho$ replaced by $X_{\theta_g^2}$.}
\begin{equation}
\label{eq:xrho}
 B^{q,(\text{id.}),C_A}_2+B^{q,\rho,(\text{nab.})}_2 = C_F C_A    \left(3 \zeta(3)-\frac{57}{8}\right) =  -\gamma_q^{(2,C_A)}+C_F b^{(C_A)}_0 X_\rho \ \ ,
 \end{equation}
 with $\gamma_q^{(2,C_A)}$ being the standard $C_F C_A$ piece of the DGLAP endpoint contribution. 
 
 \subsubsection{Relationship to NLO timelike nonsinglet DGLAP splitting kernels}
 Having extracted the $z$ dependent $\mathcal{B}^q_2$ pieces in the previous section, it is also of interest to consider the relationship between our results and the NLO timelike DGLAP splitting kernels themselves.\footnote{We note here that for the splitting kernels we intend to study in this subsection, the timelike and spacelike NLO splitting kernels have the same functional form.}
 In order to establish a formal connection with the full structure of the mass-singularity factorisation formula in QCD, we would need to integrate our results over $\rho$ or  $\theta_g^2$ and examine the resulting pole structure to recover the NLO splitting kernels. Here it is not our intention to pursue the connection from this viewpoint but rather to study the functional dependence of our results on $z$, and its possible link to the NLO splitting kernels.

 Returning to Eq.~\eqref{eq:thetadist} and focussing on its $z$ dependence, i.e. setting $\theta_g^2=1$ so as to remove the pure $\ln  \theta_g^2$ term,  we obtain a function that may be expressed as  (suppressing the overall  $\left(\frac{\alpha_s}{2\pi}\right)^2$ factor)
 \begin{multline}
 \label{eq:split}
 P^{\text{NLO},n_f}(z;\theta_g^2) = C_F  T_R n_f \left [  \frac{1+z^2}{1-z} \left(-\frac{2}{3} \ln z -\frac{10}{9}\right)-\frac{4}{3}(1-z) \right] + \\
 + C_F  T_R n_f \left[ \frac{1+z^2}{1-z} \left(\frac{2}{3} \ln (1-z)^2 +\frac{2}{3} \ln z^2\right)+\frac{2}{3}(1-z)\right].
  \end{multline}
  
  Written in this form, the top line of Eq.~\eqref{eq:split} corresponds to the $C_F T_R n_f$ piece of the {\em non-singlet} time-like splitting function $P_{qq}^{V(1)}(z)$ \cite{Ellis:1996mzs}. When integrated with a $+$ prescription on the $1/(1-z)$ factor (or equivalently after removal of terms that diverge as $z \to 1$) it gives the $-\gamma_q^{(2,n_f)}$ term of Eq.~\eqref{eq:B2int2}. The second line of Eq.~\eqref{eq:split}, when integrated with  a $+$ prescription on the $\ln(1-z)/(1-z)$ factor gives the $b_0^{(n_f)} X_{\theta_g^2}$ term of Eq.~\eqref{eq:B2int2}. For the $\rho$ distribution we can write the result in the same way in terms of $P_{qq}^{V(1)}(z)$ and a corresponding $b_0^{(n_f)} X_\rho$ piece, viz.  
  \begin{multline}
  \label{eq:split2}
  	P^{\text{NLO},n_f}(z;\rho) = C_F  T_R n_f \left [  \frac{1+z^2}{1-z} \left(-\frac{2}{3} \ln z -\frac{10}{9}\right)-\frac{4}{3}(1-z) \right] + \\
  	+ C_F  T_R n_f \left[ \frac{1+z^2}{1-z} \left(\frac{2}{3} \ln (1-z) +\frac{2}{3} \ln z\right)+\frac{2}{3}(1-z)\right] \ \ .
  \end{multline}
   The difference between the $b_0 X$ type terms for $\theta_g^2$ and $\rho$ arise, as we  stressed before, purely from the factors of $z$ and $1-z$ in the {\it{leading-order}} relation $\rho=z(1-z) \theta_g^2$.
  
  Next, let us examine the identical fermion contribution. Here there are no $b_0 X$ terms so we might expect a simple relationship of our result to the corresponding NLO timelike splitting function $P_{q\bar{q}}^{V(1)}(x)$.  That is indeed the case, however the relevant function is not the $\mathcal{P}^{\text{(id.})} (z)$ obtained in Eq.~\eqref{eq:ident}. The latter is a function of $z$, which is the energy fraction $z_3$ of parton $3$ which is one of the two identical final state quarks. It  is therefore, as we shall verify below, a contribution to the splitting function $P_{qq}^{V(1)}(x)$. If instead we carry out the calculation with fixed energy-fraction for the {\it{antiquark}}, i.e. set $z_1=x$, we obtain a different function which we expect to coincide with $P_{q\bar{q}}^{V(1)}(x)$.  To verify this we have simply performed the calculation numerically for several $x$ values and checked that the result agrees precisely with the form \cite{Curci:1980uw,Ellis:1996mzs}
  \begin{equation}
  P^{\text{(id.)}}(x) = 2 p_{qq}(-x) S_2(x) +2(1+x) \ln x +4(1-x) \ \ ,
  \end{equation}
  where $S_2(x) = -2 \mathrm{Li}_2(-x)+\frac{1}{2} \ln^2 x -2 \ln x \ln (1+x)-\frac{\pi^2}{6}.$
  The above function, on restoring the $C_F \left(C_F-\frac{C_A}{2} \right)$ factor, is just the NLO splitting kernel $P_{q\bar{q}}^{V(1)}(x)$. Moreover, the integral over $x$ of $P^{\text{(id.)}}(x)$ also gives, as expected, after supplying the colour and coupling factors,  $B_2^{q,(\text{id.)}}$ as given in Eq.~\eqref{eq:b2intid}.

  Finally we come to the pure $C_F C_A$ non-abelian term from the $q \to qgg$ splitting. The NLO collinear limit result for the $\rho$ distribution is explicitly written in Eq.~\eqref{eq:rhodistcfca}. Following a similar strategy to that for the $n_f$ piece involves setting $\rho=1$ and separating the result into two pieces, one of which yields the $b_0^{(C_A)}X_\rho$ term and the other that we may expect to be linked to the splitting kernels. It is simple to identify the terms that lead to $b_0^{(C_A)}X_\rho$. These are the same as the corresponding terms for the $n_f$ piece written on the second line of Eq.~\eqref{eq:split2} but with the replacement of $\frac{2}{3} \to -\frac{11}{6}$. The equation analogous to \eqref{eq:split2} may be written as 
  \begin{equation}
  \label{eq:splitnab}
  P^{\text{NLO},\text{nab.}}(z;\rho) = P^{\text{NLO},\text{nab.}}_{\text{sub.}}(z;\rho)-C_F C_A \left[ \frac{1+z^2}{1-z} \left(\frac{11}{6} \ln z +\frac{11}{6} \ln (1-z)\right)+\frac{11}{6}(1-z)\right],
  \end{equation}
where
 \begin{multline} 
P^{\text{NLO},\text{nab.}}_{\text{sub.}}(z;\rho)= C_F C_A \left(\frac{1+z^2}{1-z} \right)\left(\frac{11}{6} \ln z +\frac{67}{18}-\frac{\pi^2}{6} +\ln^2z+ \mathrm{Li}_2\left(\frac{z-1}{z}\right)+2 \mathrm{Li}_2(1-z)\right) \\
+C_F C_A \left(\frac{11}{6}(1-z)+\frac{3}{2} \frac{z^2 \ln z}{1-z} +\frac{1}{6}(8-5z) \right),
\end{multline}
 and where the suffix $``\text{sub.}"$ is a reminder that terms contributing to $b_0 X$ have been subtracted off to obtain this function. We can indeed use this result for $P^{\text{NLO},\text{nab.}}_{\text{sub.}}(z;\rho)$ to obtain the $C_F C_A$ piece of the splitting kernel $P_{qq}^{V(1)}(z)$. In order to do so we note that  $P_{qq}^{V(1)}(z)$  represents the splitting function where one obtains a final quark with momentum fraction $z$. In terms of all ways in which this can happen we should also consider the $C_F C_A$ contribution from the $C_F (C_F-C_A/2)$ identical particle term which yields {\it{two}} identical quarks from the splitting of an initial quark. The momentum distribution of each of these final quarks is the same and is given by the function $\mathcal{P}^{(\text{id.})}(z)$ reported in Eq.~\eqref{eq:ident}. Thus we are led to the combination
  \begin{multline}
  P^{\text{NLO},\text{nab.}}_{\text{sub.}}(z,\rho) + 2 \times \left(-C_F \frac{C_A}{2}\right) \mathcal{P}^{(\text{id.})}(z) = C_F C_A \left [\frac{1+z^2}{1-z} \left(\frac{1}{2} \ln^2 z+\frac{11}{6}\ln z+\frac{67}{18}-\frac{\pi^2}{6} \right) \right.\\ 
 \left. +(1+z) \ln z+\frac{20}{3} (1-z) \right] \ \ ,
   \end{multline} 
    where the result on the RHS does not involve any dilogarithmic terms\footnote{The following identities are useful to simplify the expression: $\mathrm{Li}_2 (1-z) =  \left(\frac{\pi^2}{6}- \ln z \ln (1-z)-\mathrm{Li}_2 (z) \right)$ and $\mathrm{Li}_2\left(\frac{z-1}{z}\right)+ \mathrm{Li}_2(1-z) =-\frac{1}{2} \ln^2 z$, with $0<z<1.$}
    and is in agreement with the $C_F C_A$ term of $P_{qq}^{V(1)}(z)$.
    
 \subsection{Combination with leading-order results}
 \label{sec:web}
In this section we consider the combination of our $\mathcal{O}\left(\alpha_s^2 \right)$ results for the  $C_F T_R n_f$ and pure $C_F C_A$ non-abelian terms (i.e. those arising from gluon branching to a pair of gluons) with the leading-order result Eq.~\eqref{eq:lo}, which will lead to the recovery of  the running coupling in the physical scheme \cite{Catani:1990rr,Banfi:2018mcq,Catani:2019rvy}, in the soft limit. Beyond the soft limit the picture that emerges remains simple to interpret and motivates a $z$ dependent extension of the soft limit result. Taking the $\theta_g^2$ distribution as an explicit example we combine the order $\alpha_s$ and order $\alpha_s^2$ results Eqs.~\eqref{eq:lo}, \eqref{eq:thetadist}, \eqref{eq:thetadistcfca},  \eqref{eq:thetadistcfca2} to define:

\begin{equation}
\left(\frac{\theta_g^2}{\sigma_0}  \frac{d^2\sigma}{d\theta_g^2 dz} \right)^{\text{tot.}} = \frac{\theta_g^2}{\sigma_0} \frac{d^2\sigma^{(1)}}{d\theta_g^2\, dz}+\left(\frac{\theta_g^2}{\sigma_0} \frac{d^2\sigma^{(2)}}{d\theta_g^2\, dz}\right)^{C_F T_R n_f}+\left(\frac{\theta_g^2}{\sigma_0} \frac{d^2\sigma^{(2)}}{d\theta_g^2\, dz}\right)^{\text{nab.}} 
\end{equation}
which gives  (recall that we have fixed $\mu_R=E$, the energy of the initial quark)
\begin{multline}
\label{eq:combined}
 \left(\frac{\theta_g^2}{\sigma_0}  \frac{d^2\sigma}{d\theta_g^2 dz} \right)^{\text{tot.}}= C_F \left(\frac{1+z^2}{1-z}\right) \left [\frac{\alpha_s\left(E^2\right)}{2\pi} +\left(\frac{\alpha_s}{2\pi} \right)^2 \left(-b_0\ln \left((1-z)^2 \theta_g^2\right)  +K \right) \right. \\
 \left. - \left(\frac{\alpha_s}{2\pi} \right)^2 b_0 \ln z  \right] + C_F b_0 \left(\frac{\alpha_s}{2\pi} \right)^2 (1-z)  + \left(\frac{\alpha_s}{2\pi} \right)^2R^{\text{nab.}}(z).
\end{multline}

We have written the result above in a form which helps to emphasise some of its main features. Firstly the two terms on the top line of Eq.~\eqref{eq:combined} combine to  produce $\alpha_s^{\text{CMW}} \left(E^2(1-z)^2\theta_g^2 \right)$ i.e.:

\begin{equation}
\label{eq:as}
  \frac{\alpha_s^{\text{CMW}} \left(E^2(1-z)^2\theta_g^2 \right)}{2\pi} \equiv  \frac{\alpha_s\left(E^2\right)}{2\pi} +\left(\frac{\alpha_s}{2\pi} \right)^2 \left(-b_0\ln \left((1-z)^2 \theta_g^2\right)  +K \right),
\end{equation}
where we have also implicitly included terms of order $\alpha_s^3$ and beyond via our change of scale for the coupling.

This agrees with the expected result in the soft limit, $z\to 1$, since the scale of the coupling  is  $E^2 (1-z)^2 \theta_g^2$ i.e. the familiar transverse momentum squared of the soft emission wrt the emitting parton and we recover the CMW prescription. On the second line we note that there are two terms proportional to the beta  function coefficient $b_0$. The presence of a $b_0 \ln z$ term is suggestive of a redefinition of the scale of the coupling in the hard collinear region to the scale $E^2z(1-z)^2\theta_g^2$. On the other hand the $b_0 (1-z)$ term is universal and originates purely in the virtual corrections, implying that it can also be absorbed into the definition of the coupling via a  $z$ dependent extension of the CMW scheme. Such a redefinition of the coupling scale and scheme would imply that the QED-like $n_f$ piece of the result is fully incorporated into the definition of the coupling, consistent with suggestions made in the past literature \cite{Dokshitzer:1995qm,Brodsky:1982gc}. One may also expect the step of defining a collinear-improved coupling to be useful from the viewpoint of inclusion of higher order (NNLL) ingredients, related to triple-collinear splitting kernels, in parton shower algorithms.\footnote{We thank Gavin Salam for several discussions related to a potential collinear improved coupling.} We shall postpone detailed discussions and definite proposals along these lines to forthcoming work \cite{GMMB}.

The function $R^{\text{nab.}}(z)$ reads
\begin{align}
	R^{\text{nab.}}(z) = C_F C_A \left[ \left(\frac{1+z^2}{1-z} \right) \left(\ln^2z+ \mathrm{Li}_2\left(\frac{z-1}{z}\right)+2\, \mathrm{Li}_2(1-z) \right) +\frac{3}{2} \frac{z^2 \ln z}{1-z} +\frac{1}{2}(2z -1) \right] \ \ , 
\end{align}
which is the remainder of the cross section solely with a $C_F C_A$ term. The above function is not soft enhanced, as the combination of terms in parenthesis multiplying $p_{qq}(z)$, vanishes as $1-z$ in the $z \to 1$ limit. The $R^{\text{nab.}}(z) $ term can be viewed as a higher-order splitting function of pure collinear origin.

We close this subsection by reminding the reader of a key property of the result Eq.~\eqref{eq:combined}. Although Eq.~\eqref{eq:combined} represents the distribution in $z$ and $\theta_g^2$, it is straightforward to 
obtain the distribution in the mass $\rho$ by integrating  Eq.~\eqref{eq:combined} over $\theta_g^2$ with the constraint $\delta(\rho-z(1-z) \theta_g^2)$ exactly as one would do in a leading-order calculation. The same holds for the distribution in any observable $v$ defined in terms of the parent gluon kinematical variables ($z,\theta_g^2$ and mass $m^2$). The fact that one can obtain one parent  gluon kinematical distribution from the other at order $\alpha_s^2$, by using a relationship valid in the limit of a massless gluon, implies  that the effect of the gluon virtuality has effectively been absorbed into the structure of  Eq.~\eqref{eq:combined}. This is reminiscent of the fact that in the soft limit and to NLL accuracy for global observables one can replace the emission of a massive gluon by a massless gluon, with the effect of the gluon branching included in the argument of the coupling and the CMW factor $K$. Therefore we may think of Eq.~\eqref{eq:combined} as an extension of the web concept beyond the soft limit and into the hard-collinear region, via an extension of the CMW coupling and a higher-order splitting function.

\subsection{$C_F^2$ abelian piece}
\label{sec:ab}
Next we examine the pure $C_F^2$ piece, i.e. the abelian $q \to q gg$ contribution. This channel is somewhat different from the gluon decay contributions we have studied thus far. Here there is no association of the results to the running coupling scale or scheme and no $b_0 X$ terms that accompany the DGLAP endpoint contribution. Accordingly in this channel one can focus purely on illustrating the extraction of $\mathcal{B}^{q,\text{(ab.)}}_2(z)$, the abelian gluon emission piece.

In Ref.~\cite{Anderle:2020mxj} we arrived at $B_2^q$ as part of the NNLL structure\footnote{We remind the reader that leading double logarithmic contributions are absent for the mMDT so that the logarithmic hierarchy starts with NLL single-logarithmic terms.} for the mMDT jet mass distribution $\frac{\rho}{\sigma_0} \frac{d\sigma}{d\rho}$, by performing an  order $\alpha_s^2$ calculation in the small $\rho$ limit.
For two real emissions, this involved considering two configurations. The first, called $\mathcal{F}^{\text{pass}}(z,\rho)$ in Ref.~\cite{Anderle:2020mxj}, had a relatively energetic large-angle emission that passes a condition $1-z_{\text{cut}}> z > z_{\text{cut}}$, with a smaller-angle emission with no constraint on its energy. Together the emissions set a value $\rho$ for the normalised jet mass squared, and taking $\rho \ll z_{\text{cut}}$ ensures that both emissions are collinear though not necessarily strongly ordered in angle. This configuration falls entirely within the jurisdiction of the triple-collinear splitting functions. 
 The second configuration, called $\mathcal{F}^{\text{fail}}(z_p,\rho)$, involves a situation where the larger-angle emission is relatively soft and fails the $\zc$ condition.\footnote{Note that $\mathcal{F}^{\text{fail}}(z_p,\rho)$ is a function of the splitting variable $z_p$, shown in Fig.~\ref{fig:qggab}} This emission is then ``groomed away'' and does not contribute to the jet mass $\rho$ which is then set by the smaller angle emission. The emission that is groomed away can be at any angle and its treatment requires going beyond collinear calculations. 
 Only the $\mathcal{F}^{\text{pass}}(z,\rho)$ term is directly related to $B^{q,\text{(ab.)}}_2$, while the $\mathcal{F}^{\text{fail}}(z_p,\rho)$ is a necessary part of recovering the full mMDT jet mass result including its leading-logarithmic terms.

Similarly for virtual corrections we needed to consider two distinct terms: firstly there is the standard  one-loop correction to the Born process ($q\bar{q}$ production) and secondly there is the one-loop correction to a $1 \to 2$ collinear splitting. Again, it is only the latter contribution that is relevant to $B^{q,\text{(ab.)}}_2$, while the former is needed as part of obtaining the full NNLL result for the  mMDT $\rho$ distribution. Performing the order $\alpha_s^2$ calculation and subtracting the NLL result, which arises entirely from the approximation of emissions strongly-ordered in angle, we obtained the relevant contribution to $B_2^{q}$, alongside NNLL ``clustering" corrections. Upon combining with the $C_F^2$ term from  the $C_F \left(C_F-C_A/2 \right)$ channel we recovered,
\begin{equation}
B_2^{q,(\text{ab.})} = -\gamma_q^{\left(2,C_F^2\right)} = C_F^2 \left(-\frac{3}{8}+\frac{\pi^2}{2}-6 \zeta(3) \right) \ \ ,
\end{equation}
though we did not study its $z$ dependence.

In this article we demonstrate more directly, by studying a simpler example, how the $\mathcal{B}^{q,\text{(ab.)}}_2(z)$ contribution may be  extracted as a difference between the triple-collinear and strongly-ordered in angle regimes, which more clearly exposes its physical origin. A key point of difference from the correlated emission calculations we have presented thus far is that when considering the decays of a parent gluon emission, requiring the parent emission to be collinear to the emitter is sufficient to constrain all three final partons to be collinear.\footnote{Recall that the correlated emission contribution vanishes exponentially in the rapidity separation of the offspring partons, which kills large angular separations between gluon offspring.}  However in the $C_F^2$ channel requiring, for instance, {\it{a small jet  mass  does not necessarily fix all partons to be collinear}}, in particular, allowing for soft wide-angle emissions unrelated to collinear physics, as already alluded to above when discussing the  $\mathcal{F}^{\text{fail}}(z,\rho)$ contribution to the mMDT jet  mass.

We shall therefore consider a collinear gluon emission (labeled $1$) with a fixed $z_1=1-z$ and angle $\theta =\theta_{13} \ll1 $ wrt the final emitted quark (see Figure~\ref{fig:qggab}). This emission will set the collinearity in the same way as the parent emission for the correlated emission channels. We shall then examine the role of more collinear real emissions  by integrating over a second emission such that $\theta_{23} < \theta$, but without imposing any other constraints. Integrating over the smaller angle emission produces divergences and accompanying finite corrections. We will then combine this calculation with the one-loop corrections to a $1\to2$ collinear splitting. It is important to note here that the above mentioned angular restriction on the secondary emission is the sole reason prohibiting an analytic calculation. Essentially, the azimuthal integral is rendered incomplete due to the angular restriction, and thus we could not provide an analytic result. Finally, we will perform the exact same calculation but using strongly-ordered dynamics with factorised $1\to 2$ splitting functions and phase space. We anticipate that the difference between the full triple-collinear limit calculation (including the one-loop correction to a $1 \to 2$ splitting) and the same calculation using the iterated $1\to 2$ kernel and phase space will suffice to directly yield $\mathcal{B}_2^{q,\text{(ab.)}}(z).$

Integrating the triple-collinear splitting function $\langle\hat{P}^{\text{(ab)}}_{g_1,g_2, q_3} \rangle$ over the region $\theta_{23}<\theta_{13}=\theta$ we obtain the following result in $4-2\epsilon$ dimensions:
\begin{equation}
\label{eq:fpass}
 \left(\frac{\theta^2}{\sigma_0} \frac{d^2\sigma}{dz d\theta^2}\right)^{\text{d-r}}  = \left(\frac{C_F \alpha_s}{2\pi}\right)^2 
 \bigg( \frac{H^{\text{soft-coll.}}(z,\theta^2,\epsilon)}{ \epsilon^2}
+ \frac{H^{\mathrm{coll.}}(z,\theta^2,\epsilon)}{\epsilon} +\frac{H^{\mathrm{soft}}(z,\theta^2,\epsilon)}{\epsilon} + H^{\text{fin.}}(z) \bigg) ,
\end{equation}
with
\begin{equation}
\label{eq:fpass2}
\begin{split}
H^{\text{soft-coll.}}(z,\theta^2,\epsilon) &= p_{qq}(z,\epsilon) z^{-4 \epsilon }(1-z)^{-2\epsilon}\theta^{-4\epsilon}
\left(1-\frac{\pi^2}{6}\epsilon^2+\mathcal{O}(\epsilon^3)\right) \ \ ,\\
H^{\text{coll.}}(z,\theta^2,\epsilon) &= p_{qq}(z,\epsilon) z^{-4 \epsilon } (1-z)^{-2\epsilon}\theta^{-4\epsilon}
\left(\frac{3}{2}+\frac{13}{2}\epsilon-\frac{2\pi^2}{3}\epsilon+\mathcal{O}(\epsilon^2)\right) \ \ ,\\
H^{\text{soft}}(z,\theta^2,\epsilon) &= 0,
\end{split}
\end{equation}
where the label $\text{d-r}$ denotes the double real contribution, $p_{qq}(z,\epsilon) =\frac{1+z^2}{1-z}-\epsilon(1-z)$, the different pole terms are labelled according to the origin of the divergence i.e. whether it comes from when the second emission is soft and collinear, pure collinear or pure soft. The contribution  $H^{\text{fin.}}(z)$ is a finite correction that we obtain via numerical integration in four dimensions.

It is to be noted that the above result is precisely the same as the $\mathcal{F}^{\text{pass}}(z,\rho)$ term we obtained for the case of fixed jet mass $\rho$ in Ref.~\cite{Anderle:2020mxj} with the  substitution $\rho \to z(1-z)\theta^2$. This relationship is just the single emission relation between $\rho$ and $\theta$ and hence exact  in  the divergent limits i.e. when the second emission is vanishingly soft ($z_2 \to 0$) or exactly collinear ($\theta_{23} \to 0$) to the final quark. The finite contribution $H^{\text{fin.}}(z)$ is identical to the mMDT jet mass case. This structure for the answer again reflects the point made previously (for the gluon decay channels), that once divergent  contributions have been removed from the triple-collinear splitting functions, the remaining finite contribution originating from relatively energetic partons, with commensurate emission angles, is independent of the observable. Accordingly one should also expect that to obtain the analogous result, from the region $\theta_{23} <\theta_{13}$, for other IRC safe observables $v$, which constrain all 3 emissions, one can use Eqs.~\eqref{eq:fpass}, \eqref{eq:fpass2} and simply replace $\theta$ by its one gluon form in terms of the observable $v$ and $z$.

\begin{figure}
\centering
\includegraphics[width=0.5 \textwidth]{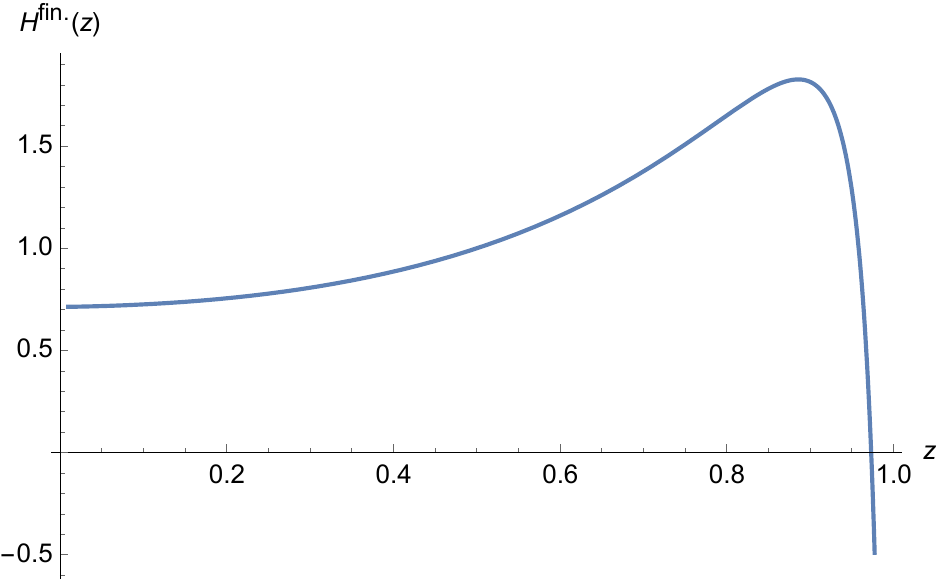}
\caption{Figure showing the dependence of $H^{\text{fin.}}(z)$ on $z$.}
\label{fig:finb2}
\end{figure}

In Ref.~\cite{Anderle:2020mxj} we evaluated the integral of $H^{\text{fin.}}(z)$, $\int_0^1 H^{\text{fin.}}(z) dz \approx 0.933\cdots$, which is consistent with the analytical form $4\zeta(3)-\frac{31}{8}$.\footnote{In  Ref.~\cite{Anderle:2020mxj} we noted that in order to recover the analytic result for $B_2^q$ we needed to identify the numerically computed value with the aforementioned analytical form. With the help of a precise numerical evaluation and the PSLQ algorithm \cite{Ferguson} it has been possible to analytically reconstruct this answer. We thank Pier Monni for his work to verify this.}
 The $z$ dependence of this function, which we have only been able to obtain numerically is shown in Fig.~\ref{fig:finb2}. It shows that as $z \to 0$ the result tends to a constant, while the steep $z \to 1$ behaviour appears consistent with the presence of a $\frac{\ln z  \ln(1-z)}{1-z}$ term. We believe, on the basis of carrying out some analytical investigation, that indeed the $\frac{\ln z  \ln(1-z)}{1-z}$ behaviour is present in the answer, but nevertheless, there might be additional terms, e.g. pure $\ln(1-z)$, that also contribute to the diverging behaviour as $z \to 1$.

\begin{figure}
\centering
\includegraphics[width=0.5 \textwidth]{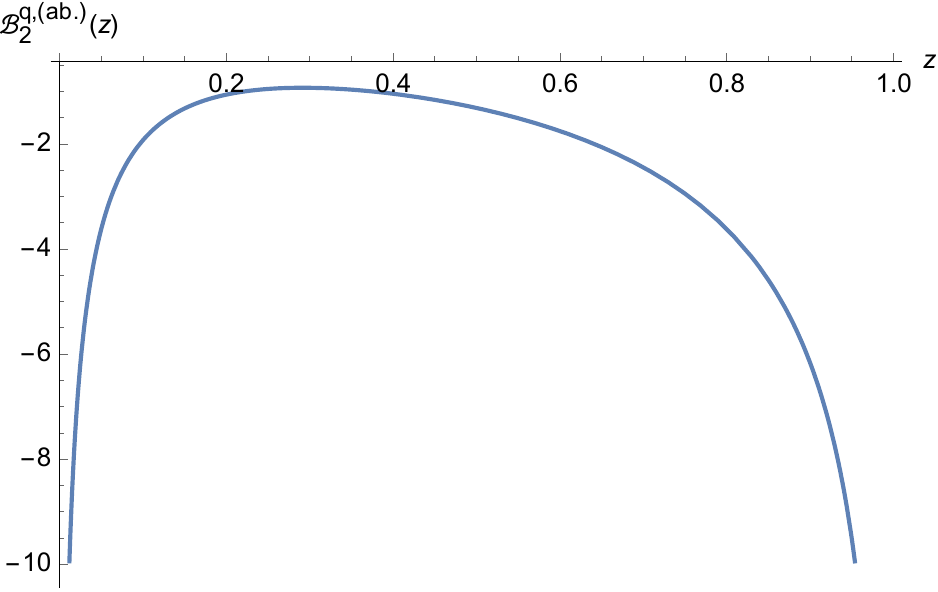}
\caption{Figure showing the dependence of $\mathcal{B}_2^{q,\text{(ab.)}}(z)$ on $z$. }
\label{fig:fullb2}
\end{figure}

In order to obtain $\mathcal{B}_2^{q,(\text{ab.})}(z)$ we will also eventually need to add to our calculation in Eqs.~\eqref{eq:fpass}, \eqref{eq:fpass2} the one-real, one-virtual contribution involving the one-loop correction to the collinear $1\to 2$ splitting given by \cite{Sborlini:2013jba}
\begin{equation}
\label{eq:one-loop}
\left(\frac{\theta^2}{\sigma_0} \frac{d^2\sigma}{dz d\theta^2}\right)^{\text{r-v}} = \left(\frac{C_F \alpha_s}{2\pi} \right)^2 \left [ p_{qq}(z,\epsilon) \left( z(1-z) \right)^{-3\epsilon} \theta^{-4\epsilon}  \left(\frac{2}{\epsilon}  \ln z+2\mathrm{Li}_2 \left(\frac{z-1}{z} \right)  \right) -1\right] \ \ .
\end{equation}

We now compute the strongly-ordered version of the double-real emission term, which we shall use as a subtraction term to generate $\mathcal{B}_2^{q,\text{(ab.)}}(z)$. This is given simply by considering a factorised product  of splitting functions and applying a factor of the single collinear emission phase-space for each emission in $4-2\epsilon$ dimensions \cite{Giele:1991vf}:
\begin{align}
\label{eq:so}
\nonumber
\left(\frac{\theta^2}{\sigma_0} \frac{d^2\sigma}{dz d\theta^2}\right)^{\text{s-o}} &= \left(\frac{C_F \alpha_s}{2\pi}\right)^2 \left(1-\frac{\pi^2}{6}\ep^2\right) p_{qq}(z,\epsilon) \left(z(1-z)  \theta^2 \right)^{-2\epsilon} \int^{\theta^2} \frac{d\theta_{23}^2}{\theta_{23}^{2(1+\epsilon)}}(z_p(1-z_p))^{-2\epsilon} p_{qq}(z_p,\epsilon)  dz_p \ \ , \\
&= \left(\frac{C_F \alpha_s}{2\pi}\right)^2 p_{qq}(z,\epsilon) \left(z(1-z)  \theta^2 \right)^{-2\epsilon} \left(\frac{1}{\ep^2} +\frac{3}{2\ep}+\frac{13}{2}-\frac{5\pi^2}{6}\right) \ \ .
\end{align}
in writing which we have expressed the result in terms of the renormalised $\overline{\text{MS}}$ coupling (with $\mu_R=E$).

We observe immediately that Eq.~\eqref{eq:so} gives an {\it{identical structure}} to that of the pole functions in the full triple-collinear result, Eqs.~\eqref{eq:fpass} \eqref{eq:fpass2}, except that the prefactor multiplying the double and single poles involves a factor $z^{-2\epsilon}$ rather than $z^{-4\epsilon}$ and there is no $H^{\text{fin.}}(z)$ term. Therefore on subtracting the strongly-ordered term from the full result we obtain:
\begin{multline}
 \left(\frac{\theta^2}{\sigma_0} \frac{d^2\sigma}{dz d\theta^2}\right)^{\text{d-r}}- \left(\frac{\theta^2}{\sigma_0} \frac{d^2\sigma}{dz d\theta^2}\right)^{\text{s-o}} \\
  = \left(\frac{C_F \alpha_s}{2\pi} \right)^2 \left(p_{qq}(z,\epsilon) \theta^{-4\epsilon}(1-z)^{-2\epsilon} \left(z^{-4\epsilon}-z^{-2\epsilon}\right) \left( \frac{1}{\epsilon^2} +\frac{3}{2\epsilon}\right)+H^{\text{fin.}}(z) \right) \ \ .
 \end{multline}

Adding in the virtual  correction we arrive at the  result

\begin{align}
 \mathcal{B}_2^{q,\text{(ab.)}}(z) &= \left(\frac{\theta^2}{\sigma_0} \frac{d^2\sigma}{dz d\theta^2}\right)^{\text{d-r}}- \left(\frac{\theta^2}{\sigma_0} \frac{d^2\sigma}{dz d\theta^2}\right)^{\text{s-o}}  +\left(\frac{\theta^2}{\sigma_0} \frac{d^2\sigma}{dz d\theta^2}\right)^{\text{r-v}}  \\
 & = \left(\frac{C_F \alpha_s}{2\pi} \right)^2 \left( \frac{1+z^2}{1-z}\left( -3 \ln z+2 \mathrm{Li}_2 \left(\frac{z-1}{z} \right) -2\ln z \ln(1-z)\right) -1+ H^{\text{fin.}}(z)\right) .
\end{align}

The functional dependence of $\mathcal{B}_2^{q,\text{(ab.)}}(z)$ on $z$ is shown in Figure \ref{fig:fullb2}. In addition to the steep behaviour as $z \to 1$, also seen for $H^{\text{fin.}}(z)$ we also note the presence of a small $z$ divergence which comes from the small $z$ approximation of $ \mathrm{Li}_2 \left(\frac{z-1}{z} \right) \approx -\frac{1}{2} \ln^2 z -\frac{\pi^2}{6}.$

Performing the integral over $z$ of $\mathcal{B}_2^{q,(\text{ab.})}(z)$, and combining with the $C_F^2$ term of the $C_F \left(C_F-C_A/2\right)$ identical particle contribution in Eq.~\eqref{eq:b2intid}, we recover the expected result:
\begin{align}
B_2^{q,\text{(ab.)}} = \left(\frac{2\pi}{\alpha_s}\right)^2 \int_0^{1}\mathcal{B}_2^{q,\text{ab.}}(z) \, dz &= \pi^2-8\zeta(3)-\frac{29}{8},  \\
B_2^{q,\text{(ab.)}}+B_2^{q,\text{(id.)},C_F^2} &= -\gamma_q^{(2,C_F^2)} \ \ ,
\end{align}
where $B_2^{q,\text{(id.)},C_F^2}$ is the $C_F^2$  term of the identical particle contribution to $B_2$ and we have used the assumed analytic form, rather than the numerical value, for the integral of $H^{\text{fin.}}(z)$. 

A few comments on the result in the abelian gluon emission channel are in order. As stated before, given the absence of any relationship to the definition of $\alpha_s$, the result obtained for  $\mathcal{B}_2^{q,\text{(ab.)}}(z)$ can be entirely viewed as an effective higher-order splitting function. While in the gluon decay channels we could relate the  $z$ dependent functions we obtained to the timelike NLO non-singlet splitting kernels, our result in the present channel {\em cannot} be directly related to the $C_F^2$ term of $P_{qq}^{V(1)}(z)$. Indeed, in our current work, for all channels we have obtained the result differential in the kinematics of the first emission, i.e the emission setting the collinearity. For the abelian gluon emission case our $z$ variable thus refers to the intermediate quark resulting from that first emission, rather than the final quark after all splittings which defines the argument of $P_{qq}^{V(1)}(z)$. Moreover we have worked with a fixed first emission and imposed a constraint on the second emission to be more collinear, which crucially impacts the structure of the result we obtain and potentially modifies the calculation relative to the one needed for $P_{qq}^{V(1)}(z)$.

\section{Conclusions}
\label{sec:concl}
In this paper we have studied collinear parton splittings at order $\alpha_s^2$ and NNLL accuracy, for splittings of a final-state quark. At our accuracy, i.e. to study NNLL collinear terms, the relevant limits of the QCD matrix-elements are given by the well-known triple-collinear splitting kernels. We have performed calculations for all relevant channels at order $\alpha_s^2$, involving both the decay of a massive gluon emitted from the initiating quark (i.e. correlated emission terms), as well as the purely abelian term with successive gluon emissions off the initiating quark. 

For the gluon decay contributions, involving $C_F T_R n_f, C_F C_A$ and  $C_F\left(C_F-C_A/2\right)$ colour channels, we have computed the distribution differential in $\rho$ and $z$, which are respectively the normalised invariant mass squared of the triple-collinear parton system and the energy fraction of the quark after emission of the massive gluon. We have also studied the distribution in $\theta_g^2$ and $z$, where $\theta_g^2$ refers to the angle made by the parent massive gluon with the final quark. In the abelian $C_F^2$ channel we have again fixed the energy fraction and angle of an initial gluon emission and then examined the role of a more collinear real emission together with the one-loop virtual correction to the initial collinear splitting.

The distributions we have derived give us several pieces of information that ought to be valuable when considering ingredients needed for an NNLL parton shower, and also of general theoretical interest. Firstly, for the gluon decay channels, in the soft limit $z \to 1$, we noted that our results produced terms that  reconstruct the correct scale of the coupling in the soft limit as well as giving the CMW coefficient $K$ \cite{Catani:1990rr}. Further, after removal of the soft and higher logarithmic order terms we obtain a pure hard-collinear term $\mathcal{B}_2^q(z)$, a differential in $z$ version of the coefficient $B_2^q$ that governs the intensity of collinear radiation at order $\alpha_s^2$.  Upon integration of $\mathcal{B}_2^q(z)$ over $z$, we recovered the standard form for $B_2^{q}$ including the $\delta(1-z)$ endpoint contributions of the timelike NLO non-singlet splitting kernels accompanied by terms of the form $b_0 X$, with $b_0$ the first perturbative coefficient of the QCD $\beta$ function. The value of $X$ is dependent on the quantity whose distribution we are considering.  We were additionally able to relate the $z$ dependent functions that emerge from our calculations to the timelike NLO non-singlet splitting kernels, after removal of terms that give rise to the $b_0 X$ contributions. 

We also discussed how our results suggest a possible extension of the scale and scheme for the soft limit CMW coupling. In particular, from our results for the $\theta_g^2$ distribution, we identified a $b_0 \ln z$ term that, when combining with the leading-order result, can be absorbed in a change of scale of the coupling, thereby modifying it relative to the known soft limit result. A further term proportional to $b_0 (1-z)$ can be considered as an extension of the CMW scheme beyond the soft limit, and doing so, alongside modifying the scale as just mentioned, completely absorbs the $n_f$ dependent pieces into the definition of the coupling. Terms that are left over are devoid of an $n_f$ term and may be viewed as effective higher-order splitting functions. We shall leave a more detailed justification and concrete proposals for the coupling scale and scheme to forthcoming work \cite{GMMB}.

For  the abelian $C_F^2$ channel we demonstrated how $\mathcal{B}_2^q(z)$ can be extracted by thinking of it as an NNLL correction that emerges naturally from the difference between the triple-collinear limit calculation (including also the one-loop correction to a $1 \to 2$ splitting) and the same calculation performed using NLL dynamics, i.e. with factorised splitting functions and phase-space. In the case of the abelian $C_F^2$ piece our results gave rise to a function which we could compute only partially analytically, while we obtained the remaining part of the result through a numerical evaluation. Again, after integrating over $z$ we recovered the known result for $B_2^q$, also in this channel.

In conclusion, as stated before, we anticipate that the findings of this paper will be of value in incorporating the NNLL hard collinear corrections, corresponding to inclusion of $B_2^{q}$, in future higher accuracy parton showers. Prior to doing so it would be necessary to extend these results to take into account collinear splittings for gluon jets and similarly account for splittings of initial state partons, which we also leave to forthcoming work.  Exploring how best to match the calculations performed here to existing shower frameworks, such as the PanScales showers \cite{Dasgupta:2020fwr,Hamilton:2020rcu,Karlberg:2021kwr}, would also still require significant further thought and effort. Nevertheless we believe that the calculations performed here and the insights we have obtained are sufficiently general so as to constitute a key step relevant to any such future work. 

\section*{Acknowledgements}
We thank Gavin Salam for a number of useful discussions through the course of this work. We also thank him, Melissa van Beekveld and Pier Monni for their valuable comments on this manuscript. We thank Jack Helliwell for discussions and collaboration on related work. We would also like to thank all our colleagues on the PanScales collaboration whose thoughts and efforts towards pushing forward the logarithmic accuracy of parton showers have directly motivated this work. This work has been funded by the European Research Council (ERC) under the European Union's Horizon 2020 research and innovation program (grant agreement No 788223) (MD and BKE) and by the U.K.'s Science and Technologies Facilities Council under grant  ST/P000800/1 (MD).

\appendix

\section{Computational details}
This appendix contains essential details about the computations presented in the manuscript. Given that the algebra is quite tedious, we restrict ourselves to present the core of the steps needed for the interested reader to reproduce the results. All intermediate steps of the computation, therefore, will be omitted.

\subsection{Recap of the {\em web} parametrization of phase space}
\label{app:web}
The web variables have been introduced in Ref.~\cite{Anderle:2020mxj} to enable the analytic computation of the distribution of the $C_F T_R n_f$ piece, {\em cf.} Eq.~\eqref{eq:rhocftf}. This parametrization of phase space appears naturally in the definition of the web function, i.e. the integrand of the soft function \cite{Banfi:2018mcq}, which is an essential ingredient in soft-gluon resummation. The extension of such parametrization in the triple-collinear limit appeared for the first time in~\cite{Anderle:2020mxj}.

The interested reader can consult Ref.~\cite{Anderle:2020mxj} for details. Here, we just collect the main result. The phase space reads:

\begin{align}\label{eq:webps}
	\text{d} \Phi^{\text{web}}_3= \frac{(4\pi)^{2\epsilon}}{256 \pi^4} \frac{2 z^{1-2\epsilon} dz}{1-z} \frac{1}{\Gamma(1-\epsilon)} \frac{d^{2-2\epsilon}k_\perp}{\Omega_{2-2\epsilon}} \frac{ds_{12}}{(s_{12})^\epsilon} \frac{dz_p}{(z_p(1-z_p))^\epsilon} \frac{1}{\Gamma(1-\epsilon)} \frac{d\Omega_{2-2\epsilon}}{\Omega_{2-2\epsilon}}  \ \ .
\end{align}
In the above, $k_\perp$ is the total transverse momentum of the parent gluon with respect to the final state quark. The energy variables, $z$ and $z_p$, are those of Figs.~(\ref{fig:triple-collinear-cfca}) and (\ref{fig:triple-collinear-fermion}), while $s_{12}$ is the invariant mass of the secondary branching, i.e. the off-shellness of the parent gluon. To explain the meaning of the solid angle $d\Omega_{2-2\epsilon}$, we first define the transverse vector
\begin{align}\label{eq:qt}
	q_\perp = \frac{k_{\perp 1}}{z_p} - \frac{k_{\perp 2}}{1-z_p}
\end{align}
and the solid angle is that of $q_\perp$, aligning $k_\perp$ along a reference axis in the transverse plane. We notice the nice feature of eq.~\eqref{eq:webps} that the double-soft limit, $z \to 1$, immediately recovers the double-soft phase space \cite{Dokshitzer:1998pt}.

\subsection{$C_F T_R n_f$: $\theta_g^2$ distribution}\label{app:cftftheta}

As we explained in the text, the fixed invariant-mass distribution has been presented in ref.~\cite{Anderle:2020mxj}. Here, we give the details of the similar computation that yields Eq.~(\ref{eq:thetadist}). The triple-collinear splitting function reads
\begin{align}
	\label{qqqprimesf}
	\la \Ph_{{\bar q}^\prime_1 q^\prime_2 q_3} \ra \, = \f{1}{2} \, 
	C_F T_R \,\f{s_{123}}{s_{12}} \left[ - \f{t_{12,3}^2}{s_{12}s_{123}}
	+\f{4z_3+(z_1-z_2)^2}{z_1+z_2} 
	+ (1-2\ep) \left(z_1+z_2-\f{s_{12}}{s_{123}}\right)
	\right] \ \ .
\end{align}
Therefore, after carrying out the azimuthal average we find
\begin{multline}
	\nonumber
	\left(\frac{\theta_g^2}{\sigma_0} \frac{d^2\sigma^{(2)}}{d\theta_g^2\, dz}\right)^{C_F T_R n_f} =  C_F T_R n_f  \left(\frac{\alpha \mu^{2\epsilon}}{2\pi} \right)^2 \left(E\, z(1-z)\theta^2_g\right)^{-2\epsilon} \int_0^1 \frac{dz_p}{\left(z_p(1-z_p)\right)^\epsilon} \int_0^\infty \frac{ds_{12}}{s_{12}^\epsilon} \\
	\times \frac{z}{s_{12}+z(1-z)^2 \theta_g^2} \bigg[ \frac{1}{s_{12}} \left( \frac{4z}{1-z} -\frac{8z z_p(1-z_p)}{(1-z)(1-\epsilon)} +(1-z)(1-2z_p)^2 +(1-2\epsilon)(1-z) \right) \\
	+\frac{1-z}{s_{12}+z(1-z)^2 \theta_g^2} \left(\frac{8zz_p(1-z_p)}{(1-z)^2}-\frac{(1+z)^2(1-2z_p)^2}{(1-z)^2}-1\right)\bigg] \ \ .
\end{multline}
Notice that the integral over the invariant mass, $s_{12}$ is convergent from above and thus can be safely extanded to infinity. In the presence of an observable, such as the jet mass $\rho$, the upper limit would be dictated by the observable, e.g. $\text{max.}\{s_{12}\}  = E^2 (1-z)\rho$. Performing the remaining integrals, and imposing charge renormalization, yields
\begin{multline}
	\left(\frac{\theta_g^2}{\sigma_0} \frac{d^2\sigma^{(2)}}{d\theta_g^2\, dz}\right)^{C_F T_R n_f} =  C_F T_R n_f  \left(\frac{\alpha_s}{2\pi} \right)^2
	\, z^{-3\epsilon} \left((1-z)^2\theta_g^2\right)^{-2\epsilon} \left(-\frac{2}{3\epsilon} p_{qq}(z,\epsilon) - \frac{10}{9} p_{qq}(z) - \frac23(1-z) \right) \ \ .
\end{multline}
Adding in the virtual corrections, i.e. Eq.~(\ref{eq:cftfvirt}), one immediately recovers Eq.~(\ref{eq:thetadist}).

\subsection{The azimuthal average}

Similar to the $C_F T_R n_f$ result first derived in Ref.~\cite{Anderle:2020mxj}, the web variables are essential to compute Eqs.~\eqref{eq:ident} and \eqref{eq:rhodistcfca}. The added complexity of the latter case arises from the azimuthal averages required in $4-2\epsilon$ dimensions. Here, we only give the core components of the calculation for any reader interested to obtain the answer themselves. We avoid any tedious, yet straightforward, steps.

The azimuthal averages we need are those of the following squared transverse vectors:
\begin{align}
	q_{\perp 1} = \frac{k_{\perp 1}}{z_p} , \quad q_{\perp 2} = \frac{k_{\perp 2}}{1-z_p} \ \ .
\end{align} 
Thus we have:
\begin{align}
	\langle \frac{1}{q_{\perp 1}^2} \rangle = \frac{\Omega_{1-2\epsilon}}{\Omega_{2-2\epsilon}} \int_0^\pi d\phi \frac{(\sin\phi)^{-2\epsilon}}{k_t^2 +(1-z_p)^2 q^2 + 2(1-z_p) q \,k_t \cos\phi} \ \ ,
\end{align}
where $q$ is defined in Eq.~\eqref{eq:qt}. Likewise, the result for $\langle 1/q_{\perp 2}^2 \rangle$ is simply obtained by replacing $z_p\to 1-z_p$. Using the simple substitution $\chi = (1+\cos \phi)/2$ we find:
\begin{align}
	\langle \frac{1}{q_{\perp 1}^2} \rangle = \frac{\pFq{2}{1}\left(1,\frac12-\epsilon,1-2 \ep; \frac{-2 x}{1-x}\right)}{\left(k_t^2 +(1-z_p)^2 q^2\right) (1-x)}, \qquad x= \frac{2(1-z_p)k_t \, q}{k_t^2 +(1-z_p)^2 q^2} \ \ .
\end{align}
Finally, one can use few basic identities of the hypergeometric function to obtain the final form:
\begin{align}\label{eq:aziav}
	\nonumber
	\langle \frac{1}{q_{\perp 1}^2} \rangle = z_p \bigg[&\frac{\Theta\left(z_p - \frac{s_{12}}{s_{12} + k_t^2}\right)}{ \left(1-\frac{(1-z_p)s_{12}}{z_p\, k_t^2}\right)^{2\ep}}  \frac{\pFq{2}{1}\left(-\ep,-2\ep,1-\ep;\frac{(1-z_p)s_{12}}{z_p \, k_t^2}\right)}{z_p \, k_t^2 -(1-z_p) s_{12}}  \\
	+& \frac{\Theta\left( \frac{s_{12}}{s_{12} + k_t^2} -z_p\right)}{ \left(1-\frac{z_p \, k_t^2}{(1-z_p) s_{12}}\right)^{2\ep}}  \frac{\pFq{2}{1}\left(-\ep,-2\ep,1-\ep;\frac{z_p \, k_t^2}{(1-z_p)s_{12}}\right)}{(1-z_p)s_{12}-z_p \, k_t^2 } \bigg]\ \ .
\end{align}

\subsection{$C_F C_A$: $\rho$ distribution}\label{app:cfcatheta}
The starting integral we need reads:
\begin{equation}
	\left(\frac{\rho}{\sigma_0} \frac{d^2\sigma^{(2)}}{d\rho\, dz}\right)^{C_F C_A} = \int \text{d}\Phi_3^{\text{web}} \frac{\left(8\pi \alpha \mu^{2\epsilon}\right)^2}{s^2_{123}} \langle \Ph^{({\rm nab.})}_{g_1 g_2 q_3} \rangle \, \rho \,\delta \left ( \rho - \frac{s_{123}}{E^2} \right) \ \ ,
\end{equation}
where 
\begin{multline}
	\langle \Ph_{g_1 g_2 q_3}^{({\rm nab})} \rangle \,
	=\Bigg\{(1-\epsilon)\left(\f{t_{12,3}^2}{4s_{12}^2}+\f{1}{4}
	-\frac{\epsilon}{2}\right)+\frac{s_{123}^2}{2s_{12}s_{13}}
	\Bigg[\frac{(1-z_3)^2(1-\epsilon)+2z_3}{z_2}\nonumber\\
	+\frac{z_2^2(1-\epsilon)+2(1-z_2)}{1-z_3}\Bigg]
	-\frac{s_{123}^2}{4s_{13}s_{23}}z_3\Bigg[\frac{(1-z_3)^2(1-\epsilon)+2z_3}{z_1z_2}
	+\epsilon(1-\epsilon)\Bigg]\nonumber\\
	+\frac{s_{123}}{2s_{12}}\Bigg[(1-\epsilon)
	\frac{z_1(2-2z_1+z_1^2) - z_2(6 -6 z_2+ z_2^2)}{z_2(1-z_3)}
	+2\epsilon\frac{z_3(z_1-2z_2)-z_2}{z_2(1-z_3)}\Bigg]\nonumber\\
	+\frac{s_{123}}{2s_{13}}\Bigg[(1-\epsilon)\frac{(1-z_2)^3
		+z_3^2-z_2}{z_2(1-z_3)}
	-\epsilon\left(\f{2(1-z_2)(z_2-z_3)}{z_2(1-z_3)}-z_1 + z_2\right) \\
	-\frac{z_3(1-z_1)+(1-z_2)^3}{z_1z_2}
	+\epsilon(1-z_2)\left(\frac{z_1^2+z_2^2}{z_1z_2}-\epsilon\right)\Bigg]\Biggr\}
	+(1\leftrightarrow 2) \ \ .
\end{multline}
It is useful to pause and comment on the collinear structure of the splitting function. Although it appears that there exists a collinear singularity in every angle, i.e. $\theta_{ij} \to 0$, in fact the only true singularity appears as $\theta_{12} \to 0$. The splitting function actually vanishes when expanded around $\theta_{13} = 0$ or $\theta_{23} = 0$, as one can easily verify. This collinear structure can be understood, from physics standpoint, as a consequence of colour coherence (angular ordering). 

 Now to the computation. The first thing to realize is that the total invariant mass is free from the azimuthal angle of the web phase space, in particular,
\begin{align}
	s_{123} = \frac{s_{12}+ zk_t^2}{1-z}  \ \ ,
\end{align}
and, therefore, the azimuthal average can be performed at the onset. The invariants $s_{13}$ and $s_{23}$ are given in terms of the squared transverse vectors, $q^2_{\perp 1}$ and $q^2_{\perp 2}$, namely
\begin{align}
	s_{13} = \frac{z z_p}{(1-z)}\, q^2_{\perp 1}, \quad s_{23} = \frac{z(1-z_p)}{(1-z) }\, q^2_{\perp 2} \ \ .
\end{align}

The full computation is somewhat tedious, so we will not proceed to give all details. Yet, we explain the necessary steps for the interested reader. First, the $\epsilon$ expansion of hypergeometric function can be performed on the integrand level and reads~\cite{Huber_2006}
\begin{align}
	\pFq{2}{1}\left(-\ep,-2\ep,1-\ep;z\right) = 1 + \epsilon^2\, \text{Li}_2(z) + \mathcal{O}\left(\epsilon^3\right) \ \ .
\end{align}
Next, to simplify things we make full use of the collinear structure of the splitting function. All terms which exhibit {\em apparent} poles in $\theta_{13}$ and $\theta_{23}$ are grouped together on the integrand level\footnote{Post the azimuthal average, these apparent poles manifest as the denominators in Eq.~(\ref{eq:aziav}) tend to zero.}. Once this is done, the remaining integrals over $s_{12}$ and $z_p$ are readily performed. The collinear pole appears as $s_{12} \to 0$, while the soft pole appears as $z_p \to 0$ or $z_p \to 1$. The final result reads, after charge renormalization,
\begin{multline}\label{eq:careal}
		\left(\frac{\rho}{\sigma_0} \frac{d^2\sigma^{(2)}}{d\rho\, dz}\right)^{C_F C_A} =  C_F C_A  \left(\frac{\alpha_s}{2\pi} \right)^2  z^{-\epsilon}(1-z)^{-2\epsilon} \rho^{-2\epsilon}\\
		\left[\frac{p_{qq}(z,\epsilon)}{\epsilon^2}
		+\left(\frac{11}{6 \epsilon}+\frac{\ln z}{\epsilon} +\frac12 \ln^2 z + \text{Li}_2\left(\frac{z-1}{z}\right)-\frac{2\pi^2}{3}\right) p_{qq}(z) +\frac{\left(z^2+4z-2\right)\ln z}{2(1-z)} + \frac{49 z^2 +45 z+40}{18(1-z)}\right] \ \ .
\end{multline}
The final ingredient is the virtual correction of the $1\to 2$ splitting which reads \cite{Sborlini:2013jba}
\begin{multline}\label{eq:virtca}
	\left(\frac{\rho}{\sigma_0} \frac{d^2\sigma_{\text{virt.}}^{(2)}}{d\rho\, dz}\right)^{C_F C_A} =  C_F C_A  \left(\frac{\alpha_s}{2\pi} \right)^2 z^{-\epsilon}(1-z)^{-\epsilon} \rho^{-2\epsilon}\, p_{qq}(z,\epsilon) \\
	\times \left[-\frac{1}{\epsilon^2} -\frac{11}{6\epsilon}\,\rho^{\epsilon} + \frac{\ln(1-z)}{\epsilon} - \frac{\ln z}{\epsilon} +\frac{2\pi^2}{3}+  \text{Li}_2\left(\frac{z}{z-1}\right)- \text{Li}_2\left(\frac{z-1}{z}\right)+\frac{1-z}{1+z^2}\right] \ \ ,
\end{multline}
and the addition of which to Eq.~(\ref{eq:careal}) immediately yields Eq.~(\ref{eq:rhodistcfca}).

\subsection{$C_F (C_F-C_A/2)$: $\rho$ distribution}\label{app:id}
Let us recall the expression of the splitting function:
\begin{align}
\label{idensf}
\langle \Ph^{({\rm id})}_{{\bar q}_1q_2q_3} \rangle \,
&= C_F \left( C_F-\f{1}{2} C_A \right)
\Biggl\{ (1-\ep)\left( \f{2s_{23}}{s_{12}} - \ep \right)\nn\\
&+ \f{s_{123}}{s_{12}}\Biggl[\f{1+z_1^2}{1-z_2}-\f{2z_2}{1-z_3}
-\ep\left(\f{(1-z_3)^2}{1-z_2}+1+z_1-\f{2z_2}{1-z_3}\right) 
- \ep^2(1-z_3)\Biggr] \nn\\
&- \f{s_{123}^2}{s_{12}s_{13}}\f{z_1}{2}\left[\f{1+z_1^2}{(1-z_2)(1-z_3)}-\ep
\left(1+2\f{1-z_2}{1-z_3}\right)
-\ep^2\right] \Biggr\} + (2\lra 3) \ \ .
\end{align}
As mentioned in the text, this colour channel contains neither soft or collinear poles, and thus one can set $\epsilon=0$ at the onset and perform the computation in 4 dimensions. The azimuthal average is performed using Eq.~(\ref{eq:aziav}), setting $\epsilon=0$, followed by the $s_{12}$ and $z_p$ integrals. The final result is given in Eq.~(\ref{eq:ident}).

\bibliographystyle{JHEP}
\bibliography{refs-md.bib}

\providecommand{\href}[2]{#2}\begingroup\raggedright\begin{thebibliography}{10}

\bibitem{KeithEllis:2019bfl}
R.~Keith~Ellis and G.~Zanderighi, \emph{{Precision QCD, higher order
  corrections and resummation in From My Vast Repertoire ...}: {Guido
  Altarelli's Legacy}},
  \href{https://doi.org/10.1142/9789813238053_0004}{\emph{World Scientific
  Singapore} (2019) 31}.

\bibitem{Azzi:2019yne}
P.~Azzi et~al., \emph{{Report from Working Group 1}: {Standard Model Physics at
  the HL-LHC and HE-LHC}},
  \href{https://doi.org/10.23731/CYRM-2019-007.1}{\emph{CERN Yellow Rep.
  Monogr.} {\bfseries 7} (2019) 1}
  [\href{https://arxiv.org/abs/1902.04070}{{\ttfamily 1902.04070}}].

\bibitem{Cepeda:2019klc}
M.~Cepeda et~al., \emph{{Report from Working Group 2}: {Higgs Physics at the
  HL-LHC and HE-LHC}},
  \href{https://doi.org/10.23731/CYRM-2019-007.221}{\emph{CERN Yellow Rep.
  Monogr.} {\bfseries 7} (2019) 221}
  [\href{https://arxiv.org/abs/1902.00134}{{\ttfamily 1902.00134}}].

\bibitem{CidVidal:2018eel}
X.~Cid~Vidal et~al., \emph{{Report from Working Group 3}: {Beyond the Standard
  Model physics at the HL-LHC and HE-LHC}},
  \href{https://doi.org/10.23731/CYRM-2019-007.585}{\emph{CERN Yellow Rep.
  Monogr.} {\bfseries 7} (2019) 585}
  [\href{https://arxiv.org/abs/1812.07831}{{\ttfamily 1812.07831}}].

\bibitem{Heinrich:2020ybq}
G.~Heinrich, \emph{{Collider Physics at the Precision Frontier}},
  \href{https://doi.org/10.1016/j.physrep.2021.03.006}{\emph{Phys. Rept.}
  {\bfseries 922} (2021) 1} [\href{https://arxiv.org/abs/2009.00516}{{\ttfamily
  2009.00516}}].

\bibitem{Banfi:2004yd}
A.~Banfi, G.~P. Salam and G.~Zanderighi, \emph{{Principles of general
  final-state resummation and automated implementation}},
  \href{https://doi.org/10.1088/1126-6708/2005/03/073}{\emph{JHEP} {\bfseries
  03} (2005) 073} [\href{https://arxiv.org/abs/hep-ph/0407286}{{\ttfamily
  hep-ph/0407286}}].

\bibitem{Catani:1991kz}
S.~Catani, G.~Turnock, B.~R. Webber and L.~Trentadue, \emph{{Thrust
  distribution in e+ e- annihilation}},
  \href{https://doi.org/10.1016/0370-2693(91)90494-B}{\emph{Phys. Lett. B}
  {\bfseries 263} (1991) 491}.

\bibitem{Banfi:2001bz}
A.~Banfi, G.~P. Salam and G.~Zanderighi, \emph{{Semi-numerical resummation of
  event shapes}},
  \href{https://doi.org/10.1088/1126-6708/2002/01/018}{\emph{JHEP} {\bfseries
  01} (2002) 018} [\href{https://arxiv.org/abs/hep-ph/0112156}{{\ttfamily
  hep-ph/0112156}}].

\bibitem{Banfi:2004nk}
A.~Banfi, G.~P. Salam and G.~Zanderighi, \emph{{Resummed event shapes at hadron
  - hadron colliders}},
  \href{https://doi.org/10.1088/1126-6708/2004/08/062}{\emph{JHEP} {\bfseries
  08} (2004) 062} [\href{https://arxiv.org/abs/hep-ph/0407287}{{\ttfamily
  hep-ph/0407287}}].

\bibitem{Bozzi:2003jy}
G.~Bozzi, S.~Catani, D.~de~Florian and M.~Grazzini, \emph{{The q(T) spectrum of
  the Higgs boson at the LHC in QCD perturbation theory}},
  \href{https://doi.org/10.1016/S0370-2693(03)00656-7}{\emph{Phys. Lett. B}
  {\bfseries 564} (2003) 65}
  [\href{https://arxiv.org/abs/hep-ph/0302104}{{\ttfamily hep-ph/0302104}}].

\bibitem{Bozzi:2005wk}
G.~Bozzi, S.~Catani, D.~de~Florian and M.~Grazzini, \emph{{Transverse-momentum
  resummation and the spectrum of the Higgs boson at the LHC}},
  \href{https://doi.org/10.1016/j.nuclphysb.2005.12.022}{\emph{Nucl. Phys. B}
  {\bfseries 737} (2006) 73}
  [\href{https://arxiv.org/abs/hep-ph/0508068}{{\ttfamily hep-ph/0508068}}].

\bibitem{Becher:2008cf}
T.~Becher and M.~D. Schwartz, \emph{{A precise determination of $\alpha_s$ from
  LEP thrust data using effective field theory}},
  \href{https://doi.org/10.1088/1126-6708/2008/07/034}{\emph{JHEP} {\bfseries
  07} (2008) 034} [\href{https://arxiv.org/abs/0803.0342}{{\ttfamily
  0803.0342}}].

\bibitem{Berger:2010xi}
C.~F. Berger, C.~Marcantonini, I.~W. Stewart, F.~J. Tackmann and W.~J.
  Waalewijn, \emph{{Higgs Production with a Central Jet Veto at NNLL+NNLO}},
  \href{https://doi.org/10.1007/JHEP04(2011)092}{\emph{JHEP} {\bfseries 04}
  (2011) 092} [\href{https://arxiv.org/abs/1012.4480}{{\ttfamily 1012.4480}}].

\bibitem{Abbate:2010xh}
R.~Abbate, M.~Fickinger, A.~H. Hoang, V.~Mateu and I.~W. Stewart, \emph{{Thrust
  at $N^{3}LL$ with Power Corrections and a Precision Global Fit for
  $\alpha_{s}(mZ)$}},
  \href{https://doi.org/10.1103/PhysRevD.83.074021}{\emph{Phys. Rev. D}
  {\bfseries 83} (2011) 074021}
  [\href{https://arxiv.org/abs/1006.3080}{{\ttfamily 1006.3080}}].

\bibitem{Becher:2010tm}
T.~Becher and M.~Neubert, \emph{{Drell-Yan Production at Small $q_T$,
  Transverse Parton Distributions and the Collinear Anomaly}},
  \href{https://doi.org/10.1140/epjc/s10052-011-1665-7}{\emph{Eur. Phys. J. C}
  {\bfseries 71} (2011) 1665}
  [\href{https://arxiv.org/abs/1007.4005}{{\ttfamily 1007.4005}}].

\bibitem{Stewart:2010pd}
I.~W. Stewart, F.~J. Tackmann and W.~J. Waalewijn, \emph{{The Beam Thrust Cross
  Section for Drell-Yan at NNLL Order}},
  \href{https://doi.org/10.1103/PhysRevLett.106.032001}{\emph{Phys. Rev. Lett.}
  {\bfseries 106} (2011) 032001}
  [\href{https://arxiv.org/abs/1005.4060}{{\ttfamily 1005.4060}}].

\bibitem{Banfi:2011dx}
A.~Banfi, M.~Dasgupta and S.~Marzani, \emph{{QCD predictions for new variables
  to study dilepton transverse momenta at hadron colliders}},
  \href{https://doi.org/10.1016/j.physletb.2011.05.028}{\emph{Phys. Lett. B}
  {\bfseries 701} (2011) 75} [\href{https://arxiv.org/abs/1102.3594}{{\ttfamily
  1102.3594}}].

\bibitem{Jouttenus:2011wh}
T.~T. Jouttenus, I.~W. Stewart, F.~J. Tackmann and W.~J. Waalewijn, \emph{{The
  Soft Function for Exclusive N-Jet Production at Hadron Colliders}},
  \href{https://doi.org/10.1103/PhysRevD.83.114030}{\emph{Phys. Rev. D}
  {\bfseries 83} (2011) 114030}
  [\href{https://arxiv.org/abs/1102.4344}{{\ttfamily 1102.4344}}].

\bibitem{Zhu:2012ts}
H.~X. Zhu, C.~S. Li, H.~T. Li, D.~Y. Shao and L.~L. Yang,
  \emph{{Transverse-momentum resummation for top-quark pairs at hadron
  colliders}},
  \href{https://doi.org/10.1103/PhysRevLett.110.082001}{\emph{Phys. Rev. Lett.}
  {\bfseries 110} (2013) 082001}
  [\href{https://arxiv.org/abs/1208.5774}{{\ttfamily 1208.5774}}].

\bibitem{Becher:2012qa}
T.~Becher and M.~Neubert, \emph{{Factorization and NNLL Resummation for Higgs
  Production with a Jet Veto}},
  \href{https://doi.org/10.1007/JHEP07(2012)108}{\emph{JHEP} {\bfseries 07}
  (2012) 108} [\href{https://arxiv.org/abs/1205.3806}{{\ttfamily 1205.3806}}].

\bibitem{Becher:2012qc}
T.~Becher and G.~Bell, \emph{{NNLL Resummation for Jet Broadening}},
  \href{https://doi.org/10.1007/JHEP11(2012)126}{\emph{JHEP} {\bfseries 11}
  (2012) 126} [\href{https://arxiv.org/abs/1210.0580}{{\ttfamily 1210.0580}}].

\bibitem{Banfi:2012jm}
A.~Banfi, P.~F. Monni, G.~P. Salam and G.~Zanderighi, \emph{{Higgs and Z-boson
  production with a jet veto}},
  \href{https://doi.org/10.1103/PhysRevLett.109.202001}{\emph{Phys. Rev. Lett.}
  {\bfseries 109} (2012) 202001}
  [\href{https://arxiv.org/abs/1206.4998}{{\ttfamily 1206.4998}}].

\bibitem{Stewart:2013faa}
I.~W. Stewart, F.~J. Tackmann, J.~R. Walsh and S.~Zuberi, \emph{{Jet $p_T$
  resummation in Higgs production at $NNLL'+NNLO$}},
  \href{https://doi.org/10.1103/PhysRevD.89.054001}{\emph{Phys. Rev. D}
  {\bfseries 89} (2014) 054001}
  [\href{https://arxiv.org/abs/1307.1808}{{\ttfamily 1307.1808}}].

\bibitem{Becher:2013xia}
T.~Becher, M.~Neubert and L.~Rothen, \emph{{Factorization and
  $N^{3}LL_{p}$+NNLO predictions for the Higgs cross section with a jet veto}},
  \href{https://doi.org/10.1007/JHEP10(2013)125}{\emph{JHEP} {\bfseries 10}
  (2013) 125} [\href{https://arxiv.org/abs/1307.0025}{{\ttfamily 1307.0025}}].

\bibitem{Kang:2013lga}
Z.-B. Kang, X.~Liu and S.~Mantry, \emph{{1-jettiness DIS event shape: NNLL+NLO
  results}}, \href{https://doi.org/10.1103/PhysRevD.90.014041}{\emph{Phys. Rev.
  D} {\bfseries 90} (2014) 014041}
  [\href{https://arxiv.org/abs/1312.0301}{{\ttfamily 1312.0301}}].

\bibitem{Kang:2013wca}
Z.-B. Kang, X.~Liu, S.~Mantry and J.-W. Qiu, \emph{{Probing nuclear dynamics in
  jet production with a global event shape}},
  \href{https://doi.org/10.1103/PhysRevD.88.074020}{\emph{Phys. Rev. D}
  {\bfseries 88} (2013) 074020}
  [\href{https://arxiv.org/abs/1303.3063}{{\ttfamily 1303.3063}}].

\bibitem{Kang:2013nha}
D.~Kang, C.~Lee and I.~W. Stewart, \emph{{Using 1-Jettiness to Measure 2 Jets
  in DIS 3 Ways}},
  \href{https://doi.org/10.1103/PhysRevD.88.054004}{\emph{Phys. Rev. D}
  {\bfseries 88} (2013) 054004}
  [\href{https://arxiv.org/abs/1303.6952}{{\ttfamily 1303.6952}}].

\bibitem{Hoang:2014wka}
A.~H. Hoang, D.~W. Kolodrubetz, V.~Mateu and I.~W. Stewart,
  \emph{{$C$-parameter distribution at N$^3$LL' including power corrections}},
  \href{https://doi.org/10.1103/PhysRevD.91.094017}{\emph{Phys. Rev. D}
  {\bfseries 91} (2015) 094017}
  [\href{https://arxiv.org/abs/1411.6633}{{\ttfamily 1411.6633}}].

\bibitem{Banfi:2014sua}
A.~Banfi, H.~McAslan, P.~F. Monni and G.~Zanderighi, \emph{{A general method
  for the resummation of event-shape distributions in $e^{+} e^{-}$
  annihilation}}, \href{https://doi.org/10.1007/JHEP05(2015)102}{\emph{JHEP}
  {\bfseries 05} (2015) 102} [\href{https://arxiv.org/abs/1412.2126}{{\ttfamily
  1412.2126}}].

\bibitem{Becher:2014aya}
T.~Becher, R.~Frederix, M.~Neubert and L.~Rothen, \emph{{Automated NNLL $+$ NLO
  resummation for jet-veto cross sections}},
  \href{https://doi.org/10.1140/epjc/s10052-015-3368-y}{\emph{Eur. Phys. J. C}
  {\bfseries 75} (2015) 154} [\href{https://arxiv.org/abs/1412.8408}{{\ttfamily
  1412.8408}}].

\bibitem{Catani:2014qha}
S.~Catani, M.~Grazzini and A.~Torre, \emph{{Transverse-momentum resummation for
  heavy-quark hadroproduction}},
  \href{https://doi.org/10.1016/j.nuclphysb.2014.11.019}{\emph{Nucl. Phys. B}
  {\bfseries 890} (2014) 518}
  [\href{https://arxiv.org/abs/1408.4564}{{\ttfamily 1408.4564}}].

\bibitem{Becher:2015lmy}
T.~Becher, X.~Garcia~i Tormo and J.~Piclum, \emph{{Next-to-next-to-leading
  logarithmic resummation for transverse thrust}},
  \href{https://doi.org/10.1103/PhysRevD.93.054038}{\emph{Phys. Rev. D}
  {\bfseries 93} (2016) 054038}
  [\href{https://arxiv.org/abs/1512.00022}{{\ttfamily 1512.00022}}].

\bibitem{Frye:2016okc}
C.~Frye, A.~J. Larkoski, M.~D. Schwartz and K.~Yan, \emph{{Precision physics
  with pile-up insensitive observables}},
  \href{https://arxiv.org/abs/1603.06375}{{\ttfamily 1603.06375}}.

\bibitem{Banfi:2016zlc}
A.~Banfi, H.~McAslan, P.~F. Monni and G.~Zanderighi, \emph{{The two-jet rate in
  $e^+e^-$ at next-to-next-to-leading-logarithmic order}},
  \href{https://doi.org/10.1103/PhysRevLett.117.172001}{\emph{Phys. Rev. Lett.}
  {\bfseries 117} (2016) 172001}
  [\href{https://arxiv.org/abs/1607.03111}{{\ttfamily 1607.03111}}].

\bibitem{Tulipant:2017ybb}
Z.~Tulip\'ant, A.~Kardos and G.~Somogyi, \emph{{Energy\textendash{}energy
  correlation in electron\textendash{}positron annihilation at NNLL + NNLO
  accuracy}}, \href{https://doi.org/10.1140/epjc/s10052-017-5320-9}{\emph{Eur.
  Phys. J. C} {\bfseries 77} (2017) 749}
  [\href{https://arxiv.org/abs/1708.04093}{{\ttfamily 1708.04093}}].

\bibitem{Bizon:2017rah}
W.~Bizon, P.~F. Monni, E.~Re, L.~Rottoli and P.~Torrielli,
  \emph{{Momentum-space resummation for transverse observables and the Higgs
  p$_{\perp}$ at N$^{3}$LL+NNLO}},
  \href{https://doi.org/10.1007/JHEP02(2018)108}{\emph{JHEP} {\bfseries 02}
  (2018) 108} [\href{https://arxiv.org/abs/1705.09127}{{\ttfamily
  1705.09127}}].

\bibitem{Bizon:2018foh}
W.~Bizo\'n, X.~Chen, A.~Gehrmann-De~Ridder, T.~Gehrmann, N.~Glover, A.~Huss
  et~al., \emph{{Fiducial distributions in Higgs and Drell-Yan production at
  N$^{3}$LL+NNLO}}, \href{https://doi.org/10.1007/JHEP12(2018)132}{\emph{JHEP}
  {\bfseries 12} (2018) 132}
  [\href{https://arxiv.org/abs/1805.05916}{{\ttfamily 1805.05916}}].

\bibitem{Procura:2018zpn}
M.~Procura, W.~J. Waalewijn and L.~Zeune, \emph{{Joint resummation of two
  angularities at next-to-next-to-leading logarithmic order}},
  \href{https://doi.org/10.1007/JHEP10(2018)098}{\emph{JHEP} {\bfseries 10}
  (2018) 098} [\href{https://arxiv.org/abs/1806.10622}{{\ttfamily
  1806.10622}}].

\bibitem{Moult:2018jzp}
I.~Moult and H.~X. Zhu, \emph{{Simplicity from Recoil: The Three-Loop Soft
  Function and Factorization for the Energy-Energy Correlation}},
  \href{https://doi.org/10.1007/JHEP08(2018)160}{\emph{JHEP} {\bfseries 08}
  (2018) 160} [\href{https://arxiv.org/abs/1801.02627}{{\ttfamily
  1801.02627}}].

\bibitem{Bell:2018mkk}
G.~Bell, B.~Dehnadi, T.~Mohrmann and R.~Rahn, \emph{{Automated Calculation of
  ${\pmb N}$-jet Soft Functions}},
  \href{https://doi.org/10.22323/1.303.0044}{\emph{PoS} {\bfseries LL2018}
  (2018) 044} [\href{https://arxiv.org/abs/1808.07427}{{\ttfamily
  1808.07427}}].

\bibitem{Chen:2018pzu}
X.~Chen, T.~Gehrmann, E.~W.~N. Glover, A.~Huss, Y.~Li, D.~Neill et~al.,
  \emph{{Precise QCD Description of the Higgs Boson Transverse Momentum
  Spectrum}}, \href{https://doi.org/10.1016/j.physletb.2018.11.037}{\emph{Phys.
  Lett. B} {\bfseries 788} (2019) 425}
  [\href{https://arxiv.org/abs/1805.00736}{{\ttfamily 1805.00736}}].

\bibitem{Banfi:2018mcq}
A.~Banfi, B.~K. El-Menoufi and P.~F. Monni, \emph{{The Sudakov radiator for jet
  observables and the soft physical coupling}},
  \href{https://doi.org/10.1007/JHEP01(2019)083}{\emph{JHEP} {\bfseries 01}
  (2019) 083} [\href{https://arxiv.org/abs/1807.11487}{{\ttfamily
  1807.11487}}].

\bibitem{Dasgupta:2021kgi}
M.~Dasgupta and J.~Helliwell, \emph{{Investigating top tagging with
  Y$_{\text{m}}$-Splitter and N-subjettiness}},
  \href{https://arxiv.org/abs/2108.09317}{{\ttfamily 2108.09317}}.

\bibitem{Buckley:2011ms}
A.~Buckley et~al., \emph{{General-purpose event generators for LHC physics}},
  \href{https://doi.org/10.1016/j.physrep.2011.03.005}{\emph{Phys. Rept.}
  {\bfseries 504} (2011) 145}
  [\href{https://arxiv.org/abs/1101.2599}{{\ttfamily 1101.2599}}].

\bibitem{Dasgupta:2018nvj}
M.~Dasgupta, F.~A. Dreyer, K.~Hamilton, P.~F. Monni and G.~P. Salam,
  \emph{{Logarithmic accuracy of parton showers: a fixed-order study}},
  \href{https://doi.org/10.1007/JHEP09(2018)033}{\emph{JHEP} {\bfseries 09}
  (2018) 033} [\href{https://arxiv.org/abs/1805.09327}{{\ttfamily
  1805.09327}}].

\bibitem{Sjostrand:2004ef}
T.~Sjostrand and P.~Z. Skands, \emph{{Transverse-momentum-ordered showers and
  interleaved multiple interactions}},
  \href{https://doi.org/10.1140/epjc/s2004-02084-y}{\emph{Eur. Phys. J. C}
  {\bfseries 39} (2005) 129}
  [\href{https://arxiv.org/abs/hep-ph/0408302}{{\ttfamily hep-ph/0408302}}].

\bibitem{Sjostrand:2014zea}
T.~Sj\"ostrand, S.~Ask, J.~R. Christiansen, R.~Corke, N.~Desai, P.~Ilten
  et~al., \emph{{An introduction to PYTHIA 8.2}},
  \href{https://doi.org/10.1016/j.cpc.2015.01.024}{\emph{Comput. Phys. Commun.}
  {\bfseries 191} (2015) 159}
  [\href{https://arxiv.org/abs/1410.3012}{{\ttfamily 1410.3012}}].

\bibitem{Banfi:2006gy}
A.~Banfi, G.~Corcella and M.~Dasgupta, \emph{{Angular ordering and parton
  showers for non-global QCD observables}},
  \href{https://doi.org/10.1088/1126-6708/2007/03/050}{\emph{JHEP} {\bfseries
  03} (2007) 050} [\href{https://arxiv.org/abs/hep-ph/0612282}{{\ttfamily
  hep-ph/0612282}}].

\bibitem{Dasgupta:2001sh}
M.~Dasgupta and G.~P. Salam, \emph{{Resummation of nonglobal QCD observables}},
  \href{https://doi.org/10.1016/S0370-2693(01)00725-0}{\emph{Phys. Lett. B}
  {\bfseries 512} (2001) 323}
  [\href{https://arxiv.org/abs/hep-ph/0104277}{{\ttfamily hep-ph/0104277}}].

\bibitem{Dasgupta:2020fwr}
M.~Dasgupta, F.~A. Dreyer, K.~Hamilton, P.~F. Monni, G.~P. Salam and G.~Soyez,
  \emph{{Parton showers beyond leading logarithmic accuracy}},
  \href{https://doi.org/10.1103/PhysRevLett.125.052002}{\emph{Phys. Rev. Lett.}
  {\bfseries 125} (2020) 052002}
  [\href{https://arxiv.org/abs/2002.11114}{{\ttfamily 2002.11114}}].

\bibitem{Hamilton:2020rcu}
K.~Hamilton, R.~Medves, G.~P. Salam, L.~Scyboz and G.~Soyez, \emph{{Colour and
  logarithmic accuracy in final-state parton showers}},
  \href{https://arxiv.org/abs/2011.10054}{{\ttfamily 2011.10054}}.

\bibitem{Karlberg:2021kwr}
A.~Karlberg, G.~P. Salam, L.~Scyboz and R.~Verheyen, \emph{{Spin correlations
  in final-state parton showers and jet observables}},
  \href{https://doi.org/10.1140/epjc/s10052-021-09378-0}{\emph{Eur. Phys. J. C}
  {\bfseries 81} (2021) 681}
  [\href{https://arxiv.org/abs/2103.16526}{{\ttfamily 2103.16526}}].

\bibitem{Forshaw:2020wrq}
J.~R. Forshaw, J.~Holguin and S.~Pl\"atzer, \emph{{Building a consistent parton
  shower}}, \href{https://doi.org/10.1007/JHEP09(2020)014}{\emph{JHEP}
  {\bfseries 09} (2020) 014}
  [\href{https://arxiv.org/abs/2003.06400}{{\ttfamily 2003.06400}}].

\bibitem{Holguin:2020joq}
J.~Holguin, J.~R. Forshaw and S.~Pl\"atzer, \emph{{Improvements on dipole
  shower colour}},
  \href{https://doi.org/10.1140/epjc/s10052-021-09145-1}{\emph{Eur. Phys. J. C}
  {\bfseries 81} (2021) 364}
  [\href{https://arxiv.org/abs/2011.15087}{{\ttfamily 2011.15087}}].

\bibitem{Nagy:2020dvz}
Z.~Nagy and D.~E. Soper, \emph{{Summations by parton showers of large
  logarithms in electron-positron annihilation}},
  \href{https://arxiv.org/abs/2011.04777}{{\ttfamily 2011.04777}}.

\bibitem{Nagy:2020rmk}
Z.~Nagy and D.~E. Soper, \emph{{Summations of large logarithms by parton
  showers}},  \href{https://arxiv.org/abs/2011.04773}{{\ttfamily 2011.04773}}.

\bibitem{Dokshitzer:1992ip}
Y.~L. Dokshitzer, G.~Marchesini and G.~Oriani, \emph{{Measuring color flows in
  hard processes: Beyond leading order}},
  \href{https://doi.org/10.1016/0550-3213(92)90211-S}{\emph{Nucl. Phys. B}
  {\bfseries 387} (1992) 675}.

\bibitem{Campbell:1997hg}
J.~M. Campbell and E.~Glover, \emph{{Double unresolved approximations to
  multiparton scattering amplitudes}},
  \href{https://doi.org/10.1016/S0550-3213(98)00295-8}{\emph{Nucl. Phys. B}
  {\bfseries 527} (1998) 264}
  [\href{https://arxiv.org/abs/hep-ph/9710255}{{\ttfamily hep-ph/9710255}}].

\bibitem{Catani:1998nv}
S.~Catani and M.~Grazzini, \emph{{Collinear factorization and splitting
  functions for next-to-next-to-leading order QCD calculations}},
  \href{https://doi.org/10.1016/S0370-2693(98)01513-5}{\emph{Phys. Lett. B}
  {\bfseries 446} (1999) 143}
  [\href{https://arxiv.org/abs/hep-ph/9810389}{{\ttfamily hep-ph/9810389}}].

\bibitem{Catani:1999ss}
S.~Catani and M.~Grazzini, \emph{{Infrared factorization of tree level QCD
  amplitudes at the next-to-next-to-leading order and beyond}},
  \href{https://doi.org/10.1016/S0550-3213(99)00778-6}{\emph{Nucl. Phys. B}
  {\bfseries 570} (2000) 287}
  [\href{https://arxiv.org/abs/hep-ph/9908523}{{\ttfamily hep-ph/9908523}}].

\bibitem{Li:2016yez}
H.~T. Li and P.~Skands, \emph{{A framework for second-order parton showers}},
  \href{https://doi.org/10.1016/j.physletb.2017.05.011}{\emph{Phys. Lett. B}
  {\bfseries 771} (2017) 59}
  [\href{https://arxiv.org/abs/1611.00013}{{\ttfamily 1611.00013}}].

\bibitem{Hoche:2017hno}
S.~H\"oche, F.~Krauss and S.~Prestel, \emph{{Implementing NLO DGLAP evolution
  in Parton Showers}},
  \href{https://doi.org/10.1007/JHEP10(2017)093}{\emph{JHEP} {\bfseries 10}
  (2017) 093} [\href{https://arxiv.org/abs/1705.00982}{{\ttfamily
  1705.00982}}].

\bibitem{Hoche:2017iem}
S.~H\"oche and S.~Prestel, \emph{{Triple collinear emissions in parton
  showers}}, \href{https://doi.org/10.1103/PhysRevD.96.074017}{\emph{Phys. Rev.
  D} {\bfseries 96} (2017) 074017}
  [\href{https://arxiv.org/abs/1705.00742}{{\ttfamily 1705.00742}}].

\bibitem{Dulat:2018vuy}
F.~Dulat, S.~H\"oche and S.~Prestel, \emph{{Leading-Color Fully Differential
  Two-Loop Soft Corrections to QCD Dipole Showers}},
  \href{https://doi.org/10.1103/PhysRevD.98.074013}{\emph{Phys. Rev. D}
  {\bfseries 98} (2018) 074013}
  [\href{https://arxiv.org/abs/1805.03757}{{\ttfamily 1805.03757}}].

\bibitem{Skrzypek:2011zw}
M.~Skrzypek, S.~Jadach, A.~Kusina, W.~Placzek, M.~Slawinska and O.~Gituliar,
  \emph{{Fully NLO Parton Shower in QCD}},
  \href{https://doi.org/10.5506/APhysPolB.42.2433}{\emph{Acta Phys. Polon. B}
  {\bfseries 42} (2011) 2433}
  [\href{https://arxiv.org/abs/1111.5368}{{\ttfamily 1111.5368}}].

\bibitem{Jadach:2016zgk}
S.~Jadach, A.~Kusina, W.~Placzek and M.~Skrzypek, \emph{{On the dependence of
  QCD splitting functions on the choice of the evolution variable}},
  \href{https://doi.org/10.1007/JHEP08(2016)092}{\emph{JHEP} {\bfseries 08}
  (2016) 092} [\href{https://arxiv.org/abs/1606.01238}{{\ttfamily
  1606.01238}}].

\bibitem{Collins:1981va}
J.~C. Collins and D.~E. Soper, \emph{{Back-To-Back Jets: Fourier Transform from
  B to K-Transverse}},
  \href{https://doi.org/10.1016/0550-3213(82)90453-9}{\emph{Nucl. Phys. B}
  {\bfseries 197} (1982) 446}.

\bibitem{Collins:1985xx}
J.~C. Collins and D.~E. Soper, \emph{{The Two Particle Inclusive Cross-section
  in $e^+ e^-$ Annihilation at {PETRA}, {PEP} and {LEP} Energies}},
  \href{https://doi.org/10.1016/0550-3213(87)90035-6}{\emph{Nucl. Phys. B}
  {\bfseries 284} (1987) 253}.

\bibitem{Kodaira:1981nh}
J.~Kodaira and L.~Trentadue, \emph{{Summing Soft Emission in QCD}},
  \href{https://doi.org/10.1016/0370-2693(82)90907-8}{\emph{Phys. Lett. B}
  {\bfseries 112} (1982) 66}.

\bibitem{Kodaira:1982az}
J.~Kodaira and L.~Trentadue, \emph{{Single Logarithm Effects in
  electron-Positron Annihilation}},
  \href{https://doi.org/10.1016/0370-2693(83)91213-3}{\emph{Phys. Lett. B}
  {\bfseries 123} (1983) 335}.

\bibitem{Moch:2004pa}
S.~Moch, J.~A.~M. Vermaseren and A.~Vogt, \emph{{The Three loop splitting
  functions in QCD: The Nonsinglet case}},
  \href{https://doi.org/10.1016/j.nuclphysb.2004.03.030}{\emph{Nucl. Phys. B}
  {\bfseries 688} (2004) 101}
  [\href{https://arxiv.org/abs/hep-ph/0403192}{{\ttfamily hep-ph/0403192}}].

\bibitem{Vogt:2004mw}
A.~Vogt, S.~Moch and J.~A.~M. Vermaseren, \emph{{The Three-loop splitting
  functions in QCD: The Singlet case}},
  \href{https://doi.org/10.1016/j.nuclphysb.2004.04.024}{\emph{Nucl. Phys. B}
  {\bfseries 691} (2004) 129}
  [\href{https://arxiv.org/abs/hep-ph/0404111}{{\ttfamily hep-ph/0404111}}].

\bibitem{Anderle:2020mxj}
D.~Anderle, M.~Dasgupta, B.~K. El-Menoufi, J.~Helliwell and M.~Guzzi,
  \emph{{Groomed jet mass as a direct probe of collinear parton dynamics}},
  \href{https://doi.org/10.1140/epjc/s10052-020-8411-y}{\emph{Eur. Phys. J. C}
  {\bfseries 80} (2020) 827}
  [\href{https://arxiv.org/abs/2007.10355}{{\ttfamily 2007.10355}}].

\bibitem{Dasgupta:2013ihk}
M.~Dasgupta, A.~Fregoso, S.~Marzani and G.~P. Salam, \emph{{Towards an
  understanding of jet substructure}},
  \href{https://doi.org/10.1007/JHEP09(2013)029}{\emph{JHEP} {\bfseries 09}
  (2013) 029} [\href{https://arxiv.org/abs/1307.0007}{{\ttfamily 1307.0007}}].

\bibitem{Larkoski:2014wba}
A.~J. Larkoski, S.~Marzani, G.~Soyez and J.~Thaler, \emph{{Soft Drop}},
  \href{https://doi.org/10.1007/JHEP05(2014)146}{\emph{JHEP} {\bfseries 05}
  (2014) 146} [\href{https://arxiv.org/abs/1402.2657}{{\ttfamily 1402.2657}}].

\bibitem{deFlorian:2001zd}
D.~de~Florian and M.~Grazzini, \emph{{The Structure of large logarithmic
  corrections at small transverse momentum in hadronic collisions}},
  \href{https://doi.org/10.1016/S0550-3213(01)00460-6}{\emph{Nucl. Phys. B}
  {\bfseries 616} (2001) 247}
  [\href{https://arxiv.org/abs/hep-ph/0108273}{{\ttfamily hep-ph/0108273}}].

\bibitem{Gatheral:1983cz}
J.~G.~M. Gatheral, \emph{{Exponentiation of Eikonal Cross-sections in
  Nonabelian Gauge Theories}},
  \href{https://doi.org/10.1016/0370-2693(83)90112-0}{\emph{Phys. Lett. B}
  {\bfseries 133} (1983) 90}.

\bibitem{Frenkel:1984pz}
J.~Frenkel and J.~C. Taylor, \emph{{Nonabelian eikonal exponentiation }},
  \href{https://doi.org/10.1016/0550-3213(84)90294-3}{\emph{Nucl. Phys. B}
  {\bfseries 246} (1984) 231}.

\bibitem{Laenen:2000ij}
E.~Laenen, G.~F. Sterman and W.~Vogelsang, \emph{{Recoil and threshold
  corrections in short distance cross-sections}},
  \href{https://doi.org/10.1103/PhysRevD.63.114018}{\emph{Phys. Rev. D}
  {\bfseries 63} (2001) 114018}
  [\href{https://arxiv.org/abs/hep-ph/0010080}{{\ttfamily hep-ph/0010080}}].

\bibitem{Dixon:2008gr}
L.~J. Dixon, L.~Magnea and G.~F. Sterman, \emph{{Universal structure of
  subleading infrared poles in gauge theory amplitudes}},
  \href{https://doi.org/10.1088/1126-6708/2008/08/022}{\emph{JHEP} {\bfseries
  08} (2008) 022} [\href{https://arxiv.org/abs/0805.3515}{{\ttfamily
  0805.3515}}].

\bibitem{deFlorian:2004mp}
D.~de~Florian and M.~Grazzini, \emph{{The Back-to-back region in e+ e-
  energy-energy correlation}},
  \href{https://doi.org/10.1016/j.nuclphysb.2004.10.051}{\emph{Nucl. Phys. B}
  {\bfseries 704} (2005) 387}
  [\href{https://arxiv.org/abs/hep-ph/0407241}{{\ttfamily hep-ph/0407241}}].

\bibitem{Davies:1984hs}
C.~T.~H. Davies and W.~J. Stirling, \emph{{Nonleading Corrections to the
  Drell-Yan Cross-Section at Small Transverse Momentum}},
  \href{https://doi.org/10.1016/0550-3213(84)90316-X}{\emph{Nucl. Phys. B}
  {\bfseries 244} (1984) 337}.

\bibitem{Catani:2000vq}
S.~Catani, D.~de~Florian and M.~Grazzini, \emph{{Universality of nonleading
  logarithmic contributions in transverse momentum distributions}},
  \href{https://doi.org/10.1016/S0550-3213(00)00617-9}{\emph{Nucl. Phys. B}
  {\bfseries 596} (2001) 299}
  [\href{https://arxiv.org/abs/hep-ph/0008184}{{\ttfamily hep-ph/0008184}}].

\bibitem{Ellis:1996mzs}
R.~K. Ellis, W.~J. Stirling and B.~R. Webber, \emph{{QCD and collider
  physics}}, vol.~8. Cambridge University Press, 2, 2011.

\bibitem{Gehrmann-DeRidder:1997fom}
A.~Gehrmann-De~Ridder and E.~W.~N. Glover, \emph{{A Complete O (alpha alpha-s)
  calculation of the photon + 1 jet rate in e+ e- annihilation}},
  \href{https://doi.org/10.1016/S0550-3213(97)00818-3}{\emph{Nucl. Phys. B}
  {\bfseries 517} (1998) 269}
  [\href{https://arxiv.org/abs/hep-ph/9707224}{{\ttfamily hep-ph/9707224}}].

\bibitem{Catani:1990rr}
S.~Catani, B.~Webber and G.~Marchesini, \emph{{QCD coherent branching and
  semiinclusive processes at large x}},
  \href{https://doi.org/10.1016/0550-3213(91)90390-J}{\emph{Nucl. Phys. B}
  {\bfseries 349} (1991) 635}.

\bibitem{Chien:2010kc}
Y.-T. Chien and M.~D. Schwartz, \emph{{Resummation of heavy jet mass and
  comparison to LEP data}},
  \href{https://doi.org/10.1007/JHEP08(2010)058}{\emph{JHEP} {\bfseries 08}
  (2010) 058} [\href{https://arxiv.org/abs/1005.1644}{{\ttfamily 1005.1644}}].

\bibitem{Curci:1980uw}
G.~Curci, W.~Furmanski and R.~Petronzio, \emph{{Evolution of Parton Densities
  Beyond Leading Order: The Nonsinglet Case}},
  \href{https://doi.org/10.1016/0550-3213(80)90003-6}{\emph{Nucl. Phys. B}
  {\bfseries 175} (1980) 27}.

\bibitem{Catani:2019rvy}
S.~Catani, D.~De~Florian and M.~Grazzini, \emph{{Soft-gluon effective coupling
  and cusp anomalous dimension}},
  \href{https://doi.org/10.1140/epjc/s10052-019-7174-9}{\emph{Eur. Phys. J. C}
  {\bfseries 79} (2019) 685}
  [\href{https://arxiv.org/abs/1904.10365}{{\ttfamily 1904.10365}}].

\bibitem{Dokshitzer:1995qm}
Y.~L. Dokshitzer, G.~Marchesini and B.~R. Webber, \emph{{Dispersive approach to
  power behaved contributions in QCD hard processes}},
  \href{https://doi.org/10.1016/0550-3213(96)00155-1}{\emph{Nucl. Phys. B}
  {\bfseries 469} (1996) 93}
  [\href{https://arxiv.org/abs/hep-ph/9512336}{{\ttfamily hep-ph/9512336}}].

\bibitem{Brodsky:1982gc}
S.~J. Brodsky, G.~P. Lepage and P.~B. Mackenzie, \emph{{On the Elimination of
  Scale Ambiguities in Perturbative Quantum Chromodynamics}},
  \href{https://doi.org/10.1103/PhysRevD.28.228}{\emph{Phys. Rev. D} {\bfseries
  28} (1983) 228}.

\bibitem{GMMB}
M.~Dasgupta, B.~K. El-Menoufi, P.~F. Monni, G.~P. Salam and M.~van Beekveld,
  \emph{{Work in Progress}}, .

\bibitem{Ferguson}
H.~R.~P. Ferguson, D.~H. Bailey and S.~Arno, \emph{{Analysis of PSLQ, an
  integer relation finding algorithm}},
  \href{https://doi.org/10.1090/S0025-5718-99-00995-3}{\emph{Math. Comp.}
  {\bfseries 68} (1999) 351}.

\bibitem{Sborlini:2013jba}
G.~F. Sborlini, D.~de~Florian and G.~Rodrigo, \emph{{Double collinear splitting
  amplitudes at next-to-leading order}},
  \href{https://doi.org/10.1007/JHEP01(2014)018}{\emph{JHEP} {\bfseries 01}
  (2014) 018} [\href{https://arxiv.org/abs/1310.6841}{{\ttfamily 1310.6841}}].

\bibitem{Giele:1991vf}
W.~T. Giele and E.~W.~N. Glover, \emph{{Higher order corrections to jet
  cross-sections in e+ e- annihilation}},
  \href{https://doi.org/10.1103/PhysRevD.46.1980}{\emph{Phys. Rev. D}
  {\bfseries 46} (1992) 1980}.

\bibitem{Dokshitzer:1998pt}
Y.~L. Dokshitzer, A.~Lucenti, G.~Marchesini and G.~P. Salam, \emph{{On the
  universality of the Milan factor for 1 / Q power corrections to jet shapes}},
  \href{https://doi.org/10.1088/1126-6708/1998/05/003}{\emph{JHEP} {\bfseries
  05} (1998) 003} [\href{https://arxiv.org/abs/hep-ph/9802381}{{\ttfamily
  hep-ph/9802381}}].

\bibitem{Huber_2006}
T.~Huber and D.~Maître, \emph{Hypexp, a mathematica package for expanding
  hypergeometric functions around integer-valued parameters},
  \href{https://doi.org/10.1016/j.cpc.2006.01.007}{\emph{Computer Physics
  Communications} {\bfseries 175} (2006) 122–144}.

\end{thebibliography}\endgroup

\end{document}